\documentclass[twocolumn,twoside,nofootinbib,showpacs,prd,aps,10pt]{revtex4}

\usepackage[hyperfootnotes=false,bookmarks=false]{hyperref}
\usepackage[dvips]{graphicx}
\usepackage[tbtags]{amsmath}
\usepackage{amssymb}

\renewcommand{\sec}[1]{Sec.~\ref{sec:#1}}

\newcommand{\subsec}[1]{Sec.~\ref{subsec:#1}}

\newcommand{\app}[1]{App.~\ref{app:#1}}

\newcommand{\eq}[1]{Eq.~\eqref{eq:#1}}
\newcommand{\eqs}[2]{Eqs.~\eqref{eq:#1} and \eqref{eq:#2}}
\newcommand{\nn}{\nonumber}

\newcommand{\Ecm}{E_\mathrm{cm}}
\newcommand{\zero}{{(0)}}
\newcommand{\one}{{(1)}}

\newcommand{\df}{\mathrm{d}}
\newcommand{\lra}{\leftrightarrow}
\newcommand{\sdt}{\!\cdot\!}
\newcommand{\bn}{\bar{n}}
\newcommand{\vn}{{\bf n}}
\newcommand{\op}{{\mathcal{O}}}
\newcommand{\bT}{\mathbf{T}}
\newcommand{\id}{\mathbf{1}}
\newcommand{\al}{\alpha}
\newcommand{\ga}{\gamma}
\newcommand{\Ga}{\Gamma}
\newcommand{\de}{\delta}
\newcommand{\eps}{\epsilon}
\newcommand{\si}{\sigma}
\newcommand{\w}{\omega}

\newcommand{\hs}{\hat{s}}
\newcommand{\hC}{\hat{C}}
\newcommand{\hO}{\hat{O}}

\begin{document}


\title{Electroweak Radiative Corrections to Higgs Production \\
via Vector Boson Fusion using SCET: Numerical Results}

\author{Fabio Siringo}
\author{Giuseppe Buccheri}
\affiliation{Dipartimento di Fisica e Astronomia, 
Universit\`a di Catania,\\
Scuola Superiore di Catania and INFN Sezione di Catania,\\
Via S.Sofia 64, I-95123 Catania, Italy.}

\date{\today}
\begin{abstract}
Electroweak radiative corrections are computed for Higgs production through vector boson fusion, 
$qq\to qqH$, which is one of the most promising channels for detecting and studying the Higgs boson 
at the LHC. Using soft-collinear effective theory, we obtain numerical results for the resummed
logarithmic contributions to the hadronic cross section 
at next-to-leading logarithmic order. We compare our results to HAWK and find good agreement below 
2 TeV where the logarithms do not dominate. The SCET method is at its best in the high LHC energy domain
where the corrections are found to be slightly larger than predicted by HAWK and by other one-loop fixed
order approximations. 
This is one of the first tests of this formalism 
at the level of a hadronic cross section, and demonstrates the viability of obtaining electroweak 
corrections for generic processes without the need for difficult
electroweak loop calculations.
\end{abstract}



\maketitle

\section{Introduction}

One of the major goals of the Large Hadron Collider (LHC), is the discovery of the Higgs boson and 
the study of its properties. The dominant production mechanism is Higgs production through gluon fusion, 
$gg\to H$. Vector boson fusion (VBF), $qq\to qqH$, is the second largest production channel and 
its cross section is about an order of magnitude smaller. Nevertheless, its measurement is very important 
for constraining the Higgs couplings and thus identifying the nature of the Higgs sector, 
see e.g.~Refs.~\cite{Duhrssen:2004cv,Zeppenfeld:2000td}. 

In this paper we will determine the electroweak corrections to the VBF process, using soft-collinear 
effective theory (SCET)~\cite{BFL,SCET1,SCET2,BPS}. The framework~\cite{Chiu:2007dg,Chiu:2009mg} for 
calculating electroweak radiative corrections in SCET allows one to obtain the logarithmic contributions 
$\al_{1,2}^n \ln^m \hs_{ij}/M_{W,Z}^2$ to the cross section, where $\hs_{ij}$ are the (partonic) generalized 
Mandelstam variables. When $\hs_{ij}$ are parametrically of the same size and 
$\hs_{ij} \gg M_{W,Z}^2$, these logarithms dominate the electroweak corrections 
and may even require resummation~\cite{Pozzorini1,Pozzorini2,Pozzorini3}. 
We will obtain numerical results at next-to-leading logarithmic (NLL) order, 
using the extension of the above framework to VBF in Ref.~\cite{Fuhrer:2010vi}, and including  
the effect of parton distribution functions (PDFs). This is one of the first tests of this formalism 
at the level of a hadronic cross section. In our knowledge only a previous test has been
reported\cite{Manohar:2012}, quite recently, 
regarding the top-quark forward-backward asymmetry in the process $q\bar q\to t\bar t$.

\begin{figure}[b] \label{fig:tree}
\centering
\includegraphics[width=0.48\textwidth]{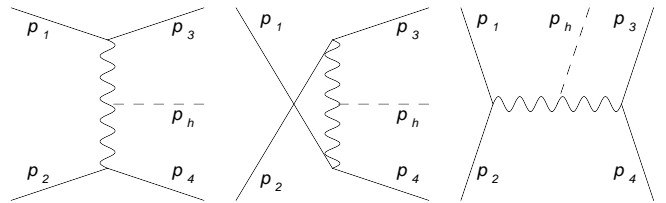}
\caption{The tree-level $t$, $u$ and $s$-channel diagrams for electroweak Higgs production. The $s$-channel process is usually referred to as Higgs strahlung rather than vector boson fusion.}
\end{figure}

The tree-level diagrams for VBF are shown in Fig.1. The two outgoing quarks produce forward jets with a 
large rapidity gap, which characterizes the VBF channel. This allows one to suppress reducible and irreducible 
(i.e.~$gg\to H+2$ jets) backgrounds using selection criteria called VBF cuts. These consist of tagging the two forward 
jets and sometimes a veto on central jets, see e.g.~Refs.~\cite{Cahn:1986zv,Barger:1994zq,Figy:2003nv,DelDuca:2006hk}. 
The VBF cuts restrict us to a region of phase space that does not satisfy the above assumptions 
on the $\hs_{ij}$ too well. We explore the validity of our results in this region by comparing with 
HAWK~\cite{Ciccolini:2007jr,Ciccolini:2007ec}, 
and find good agreement below 2 TeV. At the higher energies of LHC the method is expected to be 
more reliable, because the Mandelstam variables are quite larger than the electroweak scale and the large logarithms
dominate the electroweak corrections requiring some resummation which is here provided by SCET through the renormalization
group (RG) runnning of the coefficients in the effective interaction. A comparison with HAWK shows that in the 7-14 TeV
range of LHC the electroweak corrections are larger than predicted by fixed order perturbative approximations.
A subtle problem arises in this energy range, in some phase space regions where the Mandelstam variables
happen to be quite different in size, because the choice of the high energy scale seems to be ambiguous. The same 
problem also arises in one-loop and tree-level calculations. Our choice of an average scale seems to interpolate
smoothly between phase space regions where the logs are small and regions where they dominate the corrections.

This case study suggests that Refs.~\cite{Chiu:2009mg,Chiu:2009ft} may be used to directly obtain electroweak corrections 
for generic processes, even in regions of phase-space where they are not explicitly valid, without the need to perform 
difficult electroweak loop calculations that would have to be carried out for each individual process. 
These electroweak corrections are obtained in analytical form, and can be easily inserted in the software packages 
that have been developed for computing QCD corrections to cross sections. We will discuss in some detail how to combine 
these electroweak corrections with known QCD corrections, which is essentially multiplicative at low orders in 
perturbation theory.

QCD corrections to the VBF process have been calculated at next-to-leading order (NLO) for the inclusive cross section 
in Ref.~\cite{Han:1992hr} and for the differential cross section in Refs.~\cite{Figy:2003nv,Berger:2004pca,Arnold:2008rz}. 
The gluon-induced contribution at next-to-next-to-leading order (NNLO) was determined in Ref.~\cite{Harlander:2008xn} 
and the structure-function approach was employed in Refs.~\cite{Bolzoni:2010xr,Bolzoni:2010as,Bolzoni:2011cu} 
to get an accurate 
approximation to the full NNLO QCD corrections. The NLO electroweak corrections were determined 
in Refs.~\cite{Ciccolini:2007jr,Ciccolini:2007ec}, and are comparable in size to the QCD corrections and 
thus numerically important. 

We will now discuss in more detail how electroweak corrections are obtained in the framework 
of Ref.~\cite{Chiu:2007dg,Chiu:2009mg}. 
Here we only consider the standard model gauge group. Extensions like the minimal left-right symmetric gauge
group\cite{Siringo:2004,Siringo:2004L,Siringo:2006} will be the subject of an other paper.
The first step consists of matching onto SCET at a high scale $\mu_h \sim \sqrt{\hs_{ij}}$. 
This matching can be done in the unbroken $SU(3) \times SU(2) \times U(1)$ gauge theory, since symmetry breaking effects 
are suppressed by powers of $v/\sqrt{\hs_{ij}}$. Using the renormalization group evolution, one then runs the effective 
theory operators down to a low scale $\mu_l \sim M_Z$. At the low-scale, the $W$ and $Z$ boson are integrated out and 
one matches onto a $SU(3) \times U(1)$ effective theory, which only contains gluons and photons. 
The effects of $SU(2) \times U(1)$ symmetry breaking only enter in this low-scale matching. 

For VBF the amplitude is explicitly proportional to the vacuum expectation value (VEV), so the effective field theory 
operator is not a gauge singlet and standard resummation methods do not apply. 
Thus VBF provides a very interesting test of the method.
The extension to the VBF process 
was derived in Ref.~\cite{Fuhrer:2010vi}, and the expressions obtained there will be used through out this paper.

The paper is organized as follows: in \sec{calc} the details of the calculation are given, 
starting with the kinematics in \subsec{kin}, followed by the high-scale matching in \subsec{high}, 
the running in \subsec{run}, the low-scale matching in \subsec{low} and the electroweak cross section 
in \subsec{ew}. The necessary Passarino-Veltman functions are given in \app{PV} and the Higgs wave function and 
tadpole contribution are given in \app{RH_GH}. In \subsec{QCD}, we discuss how to combine electroweak corrections 
with known QCD corrections, but postpone the corresponding numerics to future work. Our numerical results are shown and 
discussed in \sec{results}. We conclude in \sec{concl}.

\section{Calculation}
\label{sec:calc}

\subsection{Kinematics}
\label{subsec:kin}

We will start by discussing the kinematics for the VBF process, $qq \to qqH$. The momenta of the incoming quarks (beams) 
are denoted by $p_1$ and $p_2$ and the momenta of the outgoing quarks (jets) are denoted by $p_3$ and $p_4$. 
The Higgs boson momentum is $p_h$ and we will assume that it is produced on-shell, $p_h^2 = M_h^2$. 
We will neglect quark masses, $p_1^2 = p_2^2 = p_3^2 = p_4^2 = 0$, and follow the conventions 
of Ref.~\cite{Fuhrer:2010vi} by taking all momenta to be incoming. 
The kinematic configuration is fully determined by the following generalized Mandelstam variables:
\begin{align} \label{eq:mandelst}
s&=(p_1+p_2)^2 \,, \quad
t_3=(p_1+p_3)^2 \,, \quad
t_4=(p_2+p_4)^2 \,, \nn \\
u_3&=(p_2+p_3)^2 \,, \quad
u_4=(p_1+p_4)^2 \,.
\end{align}
The other combinations can be related to these
\begin{align} \label{eq:mandelst2}
 s' = (p_3+p_4)^2 &= M_h^2 - s - t_3 - t_4 - u_3 - u_4 \,, \nn \\
 (p_1+p_h)^2 &= M_h^2 + 2p_1 \sdt p_h = M_h^2 - s - t_3 - u_4\,, \nn \\
 (p_2+p_h)^2 &= M_h^2 - s - t_4 - u_3\,, \nn \\
 (p_3+p_h)^2 &=  s + t_4 + u_4\,, \nn \\
 (p_4+p_h)^2 &=  s + t_3 + u_3\,.
\end{align}

In SCET, each collinear direction has a set of two light-cone reference vectors 
$n_i = \pm(1,\vn_i)$, $\bn_i = \pm (1,-\vn_i)$ associated with it, 
where in our conventions we take the plus (minus) sign for incoming (outgoing) particles. 
For the quarks we can simply take $n_i = \pm p_i/p_i^0$. 
For the Higgs we have to take its mass into account, $\vn_h = -{\bf p}_h/\sqrt{(p_h^0)^2 - M_h^2}$. 
The product $n_i \cdot n_j$ is then
\begin{align}
  n_i \sdt n_j &= \pm \frac{p_i \sdt p_j}{p_i^0 p_j^0}\,, \nn \\
  n_i \sdt n_h &= \pm \bigg(1- \frac{p_i \sdt p_h - p_i^0 p_h^0}{p_i^0 \sqrt{(p_h^0)^2 - M_h^2})}\bigg)\,,
\end{align}
with the plus sign when the particles are both incoming or both outgoing. 
The $p_i \cdot p_j$ and $p_i \cdot p_h$ can be directly written in terms of \eq{mandelst}, using also \eq{mandelst2}. 

At hadron colliders, such as the Tevatron and LHC, the colliding quarks ($i=1,2$) move along the beam axis and carry 
a fraction $x_i$ of the hadron momenta,
\begin{equation}
  p_i^\mu = x_i \Ecm \frac{n_i^\mu}{2}\,, \quad
  n_i^\mu = (1,0,0,\pm 1)\,, 
\end{equation}
where $\Ecm$ is the center-of-mass energy. 
In the cross section, the momentum fractions $x_i$ are integrated over. 
The distribution of momentum fractions carried by the quarks in the proton is described by PDFs. 
We parametrize the momenta of the outgoing quarks by
\begin{align} \label{eq:p_34}
  p_3 &= -E_3 (1,\sin \theta_3,0,\cos \theta_3)\,, \nn \\
  p_4 &= -E_4 (1,\sin \theta_4 \cos \varphi, \sin \theta_4 \sin \varphi,\cos \theta_4)\,, 
\end{align}
where $E_i>0$ is energy and $\theta_i$ the angle with the beam axis of particle $i$. 
The $\theta_i$ are related to (pseudo)rapidities $\eta_i$ by $\tan (\theta_i/2) = \exp(-\eta_i)$. 
The azimuthal angle between the particles is given by $\varphi$. 
Momentum conservation fixes the momentum of the Higgs and yields one additional constraint, 
as is clear from the three-body phase space,
\begin{align} \label{eq:ph_sp}
  \int \df \Phi_3 & = \int\! \prod_{i=3,4} \frac{\df^3 p_i}{(2\pi)^3 2E_i} \frac{\df^4 p_h}{(2\pi)^3} \theta(-p_h^0) \de(p_h^2 - M_h^2) \nn \\ & \quad \times
 (2\pi)^4 \de^4(p_1+p_2+p_3+p_4+p_h) \nn \\ 
  & = \frac{1}{2^6 \pi^4} \int \prod_{i=3,4} \big[E_i \df E_i\, \df (\cos \theta_i)\big] \, \df \varphi\, \theta\bigg(\sum_{j=1}^4 p_j^0\bigg) \nn \\ & \quad \times
 \de\bigg[\bigg(\sum_{j=1}^4 p_j\bigg)^2 - M_h^2\bigg] \nn \\
 & = \frac{1}{2^6 \pi^4} \int\! \df (\cos \theta_3)\, \df (\cos \theta_4)\, \df \varphi\, \df E_3\,
  \frac{E_3 f_1}{f_2^2}
\,.
\end{align}
In the first step we carry out the $p_h$ integral using the momentum conserving delta function, 
and write $p_i$ in terms of the spherical coordinates in \eq{p_34}. 
In the second step we use the on-shell condition for the Higgs to perform the $E_4$ integral, 
yielding $E_4 = f_1/f_2$ with
\begin{align}
  f_1 &= x_1 x_2 \Ecm^2 - M_h^2 - [x_1+x_2 + (x_2-x_1) \cos \theta_3] \Ecm E_3
  \nn \\
  f_2 &= [x_1+x_2 + (x_2 - x_1)\cos \theta_4] \Ecm
    -2[1-\cos \theta_3 \cos \theta_4 
  \nn \\ & \quad  
    - \cos \varphi \sin \theta_3 \sin \theta_4] E_3\,.
\end{align}
The remaining integrals in \eq{ph_sp} will be carried out numerically. 
Their boundary conditions are
\begin{align}
  &0 \leq \theta_3, \theta_4 \leq \pi
  \,, \qquad
  0 \leq \varphi \leq 2\pi
  \,, \nn \\
  &\frac{M_h^2}{\Ecm^2} \leq x_1 \leq 1
  \,, \qquad
  \frac{M_h^2}{x_1 \Ecm^2} \leq x_2 \leq 1  
  \,, \nn \\  
  &0 \leq E_3 \leq \frac{x_1 x_2 \Ecm^2 - M_h^2}{[x_1+x_2+(x_2-x_1) \cos \theta_3]\Ecm}
  \,,
\end{align}
where we included the bounds on momentum fractions of the PDFs. 
The boundaries will be further restricted by any cuts we impose on the final state. 

We can express the above variables in terms of the Mandelstam variables in \eq{mandelst},
\begin{align}
  &x_1 = \frac{\sqrt{s}}{\Ecm} e^Y\,,  \quad x_2 = \frac{\sqrt{s}}{\Ecm} e^{-Y}\,, \nn \\
  &E_3 = \!-\!\frac{1}{2\sqrt{s}}\big(t_3\, e^{-Y} \!+\! u_3\, e^Y\big), \
  E_4 = \!-\!\frac{1}{2\sqrt{s}}\big(t_4\, e^Y \!+\! u_4\, e^{-Y}\big), \nn \\
  & \cos \theta_3 = - \frac{t_3\, e^{-Y} - u_3\, e^Y}{t_3\, e^{-Y} + u_3\, e^Y}\,,\quad
  \cos \theta_4 = \frac{t_4\, e^Y - u_4\, e^{-Y}}{t_4\, e^Y + u_4\, e^{-Y}}\,, \nn \\
  & \sin \varphi = \frac{1}{\sin \theta_3\, \sin \theta_4} \Big(1-\cos\theta_3 \cos \theta_4-\frac{s'}{2E_3 E_4} \Big) \,,
\end{align}
where we also needed
\begin{equation}
  Y = \frac{1}{2} \log \frac{x_1}{x_2}\,.
\end{equation}
The rapidity $Y$ describes the boost of the partonic center-of-mass, 
which can of course not be expressed in terms of the Lorentz invariant Mandelstam variables.

For future reference, we include
\begin{equation}
 \bn_i \sdt p_i = \pm 2p_i^0\,, \quad
 \bn_h \sdt p_h = -p_h^0 + \sqrt{(p_h^0)^2 - M_h^2}\,,
\end{equation}
with a plus (minus) sign for incoming (outgoing) quarks.

\subsection{High-scale matching}
\label{subsec:high}

\subsubsection{Operator basis}

At the high scale, we can ignore the effect of electroweak symmetry breaking and work in the unbroken phase 
of $SU(2)\times U(1)$, since the partonic center-of-mass energy $\sqrt{s}$ is large compared to the VEV. 
The basis of SCET operators for vector boson fusion is given by~\cite{Fuhrer:2010vi}
\begin{align} \label{eq:op_basis}
  \op_{1A,B} &= O_1 O_A\,, O_1 O_B\,, \nn \\
  \op_{2A,B,C} &= O_2^a O_A^a\,, O_2^a O_B^a\,, O_2^a O_C^a\,, \nn \\
  \op_{3A,B} &= O_3 O_A\,, O_3 O_B\,, \nn \\
  \op_{4A,B,C} &= O_4^a O_A^a\,, O_4^a O_B^a\,, O_4^a O_C^a\,.
\end{align}
The Higgs sector is described by $O_1, \dots, O_4$
\begin{align} \label{eq:op_higgs}
O_1 &= \Phi^\dagger_h \phi_0 + \phi_0^\dagger \Phi_h\,, \nn \\
O_2^a &= \Phi^\dagger_h T^a \phi_0 - \phi_0^\dagger T^a \Phi_h\,, \nn \\
O_3 &= \Phi^\dagger_h \phi_0 - \phi_0^\dagger \Phi_h\,, \nn \\
O_4^a &= \Phi^\dagger_h T^a \phi_0 + \phi_0^\dagger T^a \Phi_h \,,
\end{align}
where $\Phi_h = W_{n_h}^\dagger \phi_{n_h}$ denotes the collinear scalar doublet $\phi_{n_h}$ that will produce the Higgs, 
plus the corresponding collinear Wilson line $W_{n_h}$. 
The field $\phi_0$ denotes a soft scalar that will attain a VEV in the broken phase. 
The basis of operators for the quarks is
\begin{align} \label{eq:op_ferm}
 O_A &= \bar \Psi_3 \ga^\mu T^a \Psi_1 \bar \Psi_4 \ga_\mu T^a \Psi_2 \,, \nn \\
 O_B &= C_F \bar \Psi_3 \ga^\mu \Psi_1 \bar \Psi_4 \ga_\mu \Psi_2 \,, \nn \\
 O^a_A &= \bar \Psi_3 \ga^\mu T^a \Psi_1 \bar \Psi_4 \ga_\mu \Psi_2 \,, \nn \\
 O^a_B &= \bar \Psi_3 \ga^\mu \Psi_1 \bar \Psi_4 \ga_\mu T^a \Psi_2 \,, \nn \\
 O^a_C &= i \eps^{abc}\, \bar \Psi_3 \ga^\mu T^b \Psi_1 \bar \Psi_4 \ga_\mu T^c\Psi_2
\,,
\end{align}
where the subscript $i=1,\dots,4$ on the field labels the particle with momentum $p_i$.  
All these fermion fields contain collinear Wilson lines $\Psi_i = W_{n_i}^\dagger \psi_{n_i}$, 
as required by collinear gauge invariance. 
In \eq{op_ferm}, $\Psi_i$ is a fermion doublet (singlet) if it is left-handed (right-handed). 
We suppressed the projectors $P_{R,L} = (1 \pm \ga_5)/2$  to keep our notation general and allow for 
both left- and right-handed quarks. 
We can consider each helicity separately and combine the contributions at the end. 
Operators $\op_A$ and $\op_C^a$ are only well-defined if all quarks are left-handed, 
and $\op_A^a$ and $\op_B^a$ are only allowed if at least one of the quarks is left-handed. 

The basis of operators in \eq{op_ferm} is not complete, 
because it assumes that the incoming particles are quarks and because it only suffices for the $t$-channel contribution. 
If one (or both) of the incoming particles is an anti-quark, we can simply obtain the corresponding basis 
and expressions by interchanging $1\lra3$ or (and) $2\lra4$. The PDFs will of course differ, and the PDF for an 
anti-quark is suppressed compared to the quark case. Similarly, to switch to the $u$-channel we interchange $3 \lra 4$. 
We can thus obtain the results for the other channels by simply making replacements and will therefore 
restrict the discussion to the $t$-channel with incoming quarks. 
Only when we square amplitudes to obtain the cross section in \subsec{ew}, will we need to be careful 
in combining the different channels. 

\subsubsection{Tree-level matching}
\label{subsec:tree}

We now perform the high-scale matching, which can be done in the unbroken phase of the electroweak gauge theory. 
Since we are working up to NLL order in the electroweak corrections, the tree-level matching suffices, 
as shown in table~\ref{tab:counting}. 
\begin{table}
  \centering
  \begin{tabular}{l | c c | c c c}
  \hline \hline
  & \multicolumn{2}{c|}{matching} & \multicolumn{3}{c}{running} \\
  & \, high scale \,& low scale \, & $\gamma$ & $\Gamma$ & $\beta$  \\ \hline
  LO & $0$-loop & - & - & - & $1$-loop \\
  NLO & $1$-loop & - & - & - & $2$-loop \\ \hline
  LL & $0$-loop & $0$-loop & - & $1$-loop & $1$-loop\\  
  NLL & $0$-loop & $1$-loop & \,$1$-loop & $2$-loop & $2$-loop\\
  NNLL & $1$-loop & $2$-loop & \, $2$-loop & $3$-loop & $3$-loop\\
  \hline\hline
  \end{tabular}
\caption{Order counting in fixed-order and resummed perturbation theory. $\Ga$ and $\ga$ denote the 
cusp and non-cusp part of the anomalous dimension, and $\beta$ denotes the beta function.}
\label{tab:counting}
\end{table}
The relevant terms in the Lagrangian that couple the scalar doublet to the gauge fields are 
\begin{align}
{\cal L}=&\,\frac{1}{4} \big(g_2^2 A^a_\mu A^{a\mu} + g_1^2 B_\mu B^\mu \big)\Phi^\dagger\Phi \nn \\
& + g_1g_2 B^\mu A^a_\mu\, \Phi^\dagger T^a \Phi \,.
\end{align}
The full-theory diagrams were shown in Fig.1. Depending on the parton types, 
only some of the diagrams contribute. 
Phenomenologically, vector boson fusion is most interesting when the jets are in the forward direction, 
for which the $s$-channel contribution is suppressed. We therefore do not include the s-channel.
Matching the $t$-channel diagram for incoming quarks onto the operators in \eq{op_basis}, yields
\begin{align} \label{eq:c_def}
  C_{1A}(\mu) &= \frac{i g_2^4(\mu)}{2t_3 t_4}\, \eta_1 \eta_2\,, \ \
  C_{1B}(\mu) = \frac{2i g_1^4(\mu)}{3t_3 t_4}\, Y_1 Y_2\,, \nn \\
  C_{4A}(\mu) &= \frac{i g_1^2(\mu) g_2^2(\mu)}{t_3 t_4}\, \eta_1 Y_2\,, \nn \\
  C_{4B}(\mu) &= \frac{i g_1^2(\mu) g_2^2(\mu)}{t_3 t_4}\, Y_1 \eta_2\,. 
\end{align}
The other Wilson coefficients vanish at tree level. Here, the variable $\eta_i$, 
is defined as $\eta_i=1$ if the particle with label $i$ is left-handed and $\eta_i=0$ if the particle is right-handed. 
The hypercharge $Y=1/6$ for left-doublets, $Y=2/3$ for right-handed up-type quarks and $Y=-1/3$ 
for right-handed down-type quarks. 
The contribution for the $u$-channel and $s$-channel can be obtained through a permutation, as discussed in \subsec{ew}.

\subsection{Running}
\label{subsec:run}

The running of the Wilson coefficients is described by the RG equation 
\begin{equation} \label{eq:RGE}
 \mu \frac{\df}{\df\mu}C(\mu) = \ga(\mu) C(\mu)
\,,\end{equation}
where $\ga$ is the anomalous dimension for the operators. It should be noted that this is a matrix equation, 
corresponding to the 10 operators in \eq{op_basis}. In Ref.~\cite{Fuhrer:2010vi}, the anomalous dimension was derived, 
\begin{align} \label{eq:ga_total}
 \ga &= \tilde\ga + \hat \ga \nn \\
 \hat \ga & =
  \frac{\al}{4\pi} \bigg[ 
 \bT_h \sdt \bT_0 (4 L_h + 2)  
 + \sum_i \bT_i \sdt \bT_0\, 4 L_i \bigg] 
\,.
\end{align}
The $\tilde \gamma$ denotes the anomalous dimension one would obtain by summing over the soft and collinear 
functions~\cite{Chiu:2009mg,Chiu:2009ft} for all the external particles 
(i.e.~the quarks and Higgs, but not the VEV), and $\hat \gamma$ describes a new contribution. 
In \eq{ga_total}, the sum on $i$ runs over the quarks and we introduce the shorthand 
\begin{equation}
L_i = \log \frac{\bn_i \cdot p_i}{\mu}
\,.
\end{equation}
The gauge group generator $\bT_i$ acts on particle $i$ as,
\begin{align}
\big(\mathbf{T}^a_i \Psi_j\big)_\alpha &= -T^a_{\alpha \beta} \Psi_{j\beta}\, \delta_{ij}\,, \nn \\
\big(\mathbf{T}^a_i \bar \Psi_j\big)_\alpha &= \bar \Psi_{j\beta} T^a_{\beta \alpha}\, \delta_{ij}\,.
\end{align}
The anomalous dimension was derived in the unbroken phase of the gauge theory, hence the presence of the 
generator $\bT_0$ that acts on the soft field $\phi_0$ (in the broken phase $\phi_0$ gets a VEV). 

We now give the explicit expressions, needed for our numerical analysis. Under $SU(2)$, the ten operators 
in \eq{op_basis} split into subsets $\{\op_{1A},\op_{1B},\op_{2A},\op_{2B},\op_{2C}\}$ 
and $\{\op_{3A},\op_{3B},\op_{4A},\op_{4B},\op_{4C}\}$, 
that do not mix under renormalization. 
In these bases, the $SU(2)$ anomalous dimension is identical for both subsets and will be given 
in terms of $5\times5$ matrices. The $SU(2)$ soft function is
\begin{widetext}
\begin{align} \label{eq:ga_S}
  \ga_S^{SU(2)} = 
  \Ga(\al_2) \bigg[
  \frac{3}{4} c_1 \id_{5\times5} + \frac{1}{4}
  \begin{pmatrix}
    - 4c_1 + 2c_3 & 3 c_2 & c_{2-} & c_{1-} & c_{1+} - c_{2+} \\
    c_2 & 0 & c_{1-} & c_{2-} & 0 \\
    c_{2-} & 3c_{1-} & -4 \eta_1 U_{13} + 2c_{1+} & c_2 & c_{3+} - c_{2-} \\
    c_{1-} & 3c_{2-} & c_2 & -4 \eta_2 U_{24} +2c_{2+} & -c_{3-} + c_{1-} \\
    2(c_{1+} - c_{2+}) & 0 & 2 (c_{3+} - c_{2-}) & -2 (c_{3-} - c_{1-}) & - 4c_1 + c_3 + c_{1+} + c_{2+}
  \end{pmatrix} \bigg]
   \,,
\end{align}
\end{widetext}
where
\begin{equation}
 U_{ij} = \log \frac{-n_i \sdt n_j - i0}{2}\,,
\end{equation}
and
\begin{align} \label{eq:c_def2}
 c_1 &= \eta_1 U_{13} + \eta_2 U_{24}\,, \nn \\
 c_2 &= \eta_1 \eta_2 (U_{14} + U_{23} - U_{12} -U_{34})\,, \nn \\
 c_3 &= \eta_1 \eta_2 (U_{12} + U_{14} + U_{23} + U_{34})\,, \nn \\
 c_{1\pm} &= \eta_1 (U_{1h} \pm U_{3h})\,, \nn \\
 c_{2\pm} &= \eta_2 (U_{2h} \pm U_{4h})\,, \nn \\
 c_{3\pm} &= \eta_1 \eta_2 [(U_{12} - U_{34}) \pm (U_{23} - U_{14})]\,.
\end{align}
Note that if only one of the $\eta_i$ is zero, \eq{ga_S} contains entries for operators that are not 
allowed when one of the fields is an $SU(2)$ singlet, e.g.~$c_1$ in the upper-left corner. 
However, they are harmless since the Wilson coefficients for these operators vanish and these entries 
do not lead to mixing with any of the allowed operators.
The cusp anomalous dimension $\Ga(\al)$~\cite{Korchemsky:1987wg} is known up to three loops~\cite{Moch:2004pa}. 
We only need the two-loop result
\begin{align} \label{eq:ga_cusp}
  \Ga(\al) &= \frac{\al}{\pi}\bigg[1 + \sum_n \Big(\frac{\al}{4\pi}\Big)^n K^{(n)}\bigg]\,, \nn \\  
  K^\one &= \Big(\frac{67}{9} - \frac{\pi^2}{3}\Big)C_A - \frac{20}{9} n_F T_F 
  - \frac{8}{9} n_S T_S\,. 
\end{align}
We obtained the scalar contribution from the fermion case, since both only enter through the vacuum polarization 
at this order. For $SU(2)$, $n_F$ and $n_S$ are the number of fermion and scalar doublets. 
For $U(1)_Y$, $n_F T_F$ and $n_S T_S$ get replaced by the sum over the squared hypercharges $Y$ of the particles.
Explicitly, the group theory constants are given by
\begin{align}
SU(2)\!:& \ C_A = 2\,, T_F = T_S = \frac{1}{2}\,, n_F = 6\,, n_S = 1\,, \nn \\
U(1)_Y\!:& \ C_A=0\,, n_F T_F \to 5\,, n_S T_S \to \frac{1}{2}\,.
\end{align}

We now move on to the $U(1)_Y$ soft function, which mixes the Higgs operators in \eq{op_higgs} but not the 
quark operators in \eq{op_ferm}. 
This mixing takes place in the subsets $\{O_1, O_3\}$ and $\{O_2, O_4\}$ of \eq{op_higgs}, 
and the anomalous dimension for each subset is identical. In this basis the $U(1)_Y$ soft function is given by
\begin{align} \label{eq:ga_S1}
  \ga_S^{U(1)_Y} = \Ga(\al_1)\bigg[c_4 + \frac{c_5}{2}
  \begin{pmatrix}
   0 & 1 \\
   1 & 0
  \end{pmatrix} \bigg]
\end{align}
where
\begin{align}
  c_4 &= Y_1^2 U_{13} + Y_2^2 U_{24} + Y_1 Y_2 (U_{14} + U_{23} - U_{12} - U_{34})\,, \nn \\
  c_5 & = Y_1 (U_{1h} - U_{3h}) + Y_2 (U_{2h} - U_{4h})\,.
\end{align}

The collinear functions for the quarks and for the Higgs are given by
\begin{align}
  \ga_C^i &= \Big[\frac{3}{4} \Ga(\al_2) \eta_i + \Ga(\al_1) Y_i^2\Big] L_i - \frac{9 \al_2}{16\pi} \eta_i - \frac{3 \al_1}{4\pi} Y_i^2\,, \nn \\
  \ga_C^h &= \Big[\frac{3}{4} \Ga(\al_2) + \frac{1}{4} \Ga(\al_1) \Big] L_h - \frac{3 \al_2}{4\pi} - \frac{\al_1}{4\pi} + \frac{3y_t^2}{16\pi^2}\,,
\end{align}
where we included the contribution from the top Yukawa $y_t$ to the Higgs wave function renormalization.

The new contribution in \eq{ga_total} is given by~\cite{Fuhrer:2010vi}
\begin{widetext}
\begin{align} \label{eq:ga_x}
  \hat \ga^{SU(2)} &= 
 \frac{1}{4}
 \Big[\Ga(\al_2) L_h + \frac{\al_2}{2\pi}\Big]
 \begin{pmatrix}
  -3 & 0 & 0 & 0 & 0 \\
  0 & -3 & 0 & 0 & 0\\
  0 & 0 & 1 & 0 & 0\\
  0 & 0 & 0 & 1 & 0\\
  0 & 0 & 0 & 0 & 1
 \end{pmatrix} 
  +
  \frac{1}{4}  \Ga(\al_2)
  \begin{pmatrix}
    0 & 0 & \eta_2 L_{2/4} & \eta_1 L_{1/3} & \eta_1 \eta_2 L_{13/24} \\
    0 & 0 & \eta_1 L_{1/3} & \eta_2 L_{2/4} & 0 \\
    \eta_2 L_{2/4} & 3\eta_1 L_{1/3} & -2 \eta_1 L_{13} & 0 & \eta_2 L_{2/4} \\
    \eta_1 L_{1/3} & 3\eta_2 L_{2/4} & 0 & -2\eta_2 L_{24} & \eta_1 L_{3/1} \\
    2\eta_1 \eta_2 L_{13/24} & 0 & 2 \eta_2 L_{2/4} & 2 \eta_1 L_{3/1} & - \eta_1 \eta_2 L_{1234}
  \end{pmatrix} 
   \,, \nn \\
  \hat \ga^{U(1)_Y} &= - \frac{1}{4}\Big[\Ga(\al_1) L_h + \frac{\al_1}{2\pi}\Big] \id_{2\times2} 
 + \frac{1}{2} \Ga(\al_1) (Y_1 L_{1/3}+ Y_2 L_{2/4})
  \begin{pmatrix}
   0 & 1 \\
   1 & 0
  \end{pmatrix}\,, \nn \\
 \hat \ga^\lambda &= \frac{c^\op\lambda}{4\pi^2} \,,
\end{align}
\end{widetext}
where we have used the same bases as for the soft anomalous dimensions and abbreviated 
\begin{equation}
L_{ab\dots/cd\dots} = (L_a + L_b + \dots) - (L_c + L_d + \dots)\,.
\end{equation}
The last line in \eq{ga_x} is the Higgs rescattering contribution, 
with coefficient $c^\op = \{3,0,0,1\}$ for the operator $\{O_1, O_2^a, O_3, O_4^a\}$. 

The total anomalous dimension in \eq{ga_total} is then
\begin{equation} \label{eq:ga_result}
\ga = \ga_S^{SU(2)} + \ga_S^{U(1)} + \sum_i \ga_C^i + \hat \ga^{SU(2)} + \hat \ga^{U(1)_Y} + \hat \ga^\lambda\,,
\end{equation}
where each of the ingredients has to be appropriately written as a $10\times10$ matrix and the sum on $i$ 
runs over the quarks and the Higgs. 
A simple crosscheck on this expression comes from reparametrization invariance (RPI)~\cite{Chay:2002vy, Manohar:2002fd}. 
Under the RPI-III transformation, 
\begin{equation} \label{eq:RPI}
n_i^\mu \to e^{\kappa_i} n_i^\mu
\,,\qquad
\bn_i^\mu \to e^{-\kappa_i} \bn_i^\mu
\,,
\end{equation}
where $\kappa_i$ is a parameter that can be chosen different for each collinear sector. 
The individual ingredients in \eq{ga_result} are not RPI-III invariant but we find that their sum is invariant, 
as should be the case. 

\subsection{Low-scale matching}
\label{subsec:low}

\subsubsection{Operator basis}

At low energies the effects of electroweak symmetry breaking need to be taken into account. 
In this section we will match onto a basis of operators in the broken phase of the gauge group. 
The Higgs part of the operators gets matched onto a $v(\mu_l) h_n$, where we include the running of the VEV. 
The quark part in \eq{op_ferm} can be matched onto the set
\begin{align} \label{eq:br_basis}
  \hO_A = \bar u_3 \ga^\mu u_1 \bar u_4 \ga_\mu u_2 \,, \nn \\
  \hO_B = \bar u_3 \ga^\mu u_1 \bar d_4 \ga_\mu d_2 \,, \nn \\
  \hO_C = \bar d_3 \ga^\mu d_1 \bar u_4 \ga_\mu u_2 \,, \nn \\
  \hO_D = \bar d_3 \ga^\mu d_1 \bar d_4 \ga_\mu d_2 \,, \nn \\
  \hO_E = \bar d_3 \ga^\mu u_1 \bar u_4 \ga_\mu d_2 \,, \nn \\
  \hO_F = \bar u_3 \ga^\mu d_1 \bar d_4 \ga_\mu u_2 \,.
\end{align}
Here $u$ and $d$ denote up and down-type fields. 
For operators $\hO_A,\dots, \hO_D$, each pair of fields $\bar \psi \ga^\mu \psi$ is a pair of left-handed or 
right-handed fields, whereas in $\hO_E$ and $\hO_F$ all fields are left handed.

\subsubsection{Tree-level matching}

We first perform the tree-level matching for the Higgs part of the operator, given in \eq{op_higgs}. 
At tree level, the soft scalar field simply attains a VEV and the collinear scalar field produces a Higgs, 
leading to
\begin{equation} \label{eq:Higgs_tree}
  \{O_1,O_2^a,O_3,O_4^a\} \to \{1,0,0,-\de^{a3}/2\} v h_n\,.
\end{equation}
We now match \{$\op_{1A,B}$, $\op_{4A,B,C}$\} onto the quark operators $\{\hO_A,\dots, \hO_F\}$ in \eq{br_basis}, 
ignoring $O_2^a$ and $O_3$ which vanish at tree level. This is described by a $6 \times 5$ matching matrix $R^\zero$,
\begin{align}
  O_i \to v h_n \sum_J \hO_j R^\zero_{ji}\,, \quad
  \hC_j = v\sum_i R^\zero_{ji}C_i 
\,,\end{align}
such that
\begin{equation}
  \sum_i C_i O_i \to \sum_i \hC_i h_n \hO_i 
\,.\end{equation}
The matrix $R^\zero$ is given by
\begin{equation}
R^\zero = \frac{1}{4}
\begin{pmatrix}
1 & 3c_{u,1} c_{u,2} & -c_{u,2} & -c_{u,1} & 0\\
-1 & 3c_{u,1} c_{d,2} & -c_{d,2} & c_{u,1} & 0\\
-1 & 3c_{d,1} c_{u,2} & c_{u,2} & -c_{d,1} & 0\\
1 & 3c_{d,1} c_{d,2} & c_{d,2} & c_{d,1} & 0\\
2 & 0 & 0 & 0 & -1 \\
2 & 0 & 0 & 0 & 1
\end{pmatrix}\,.
\end{equation}
Some of the quark fields in the unbroken basis in \eq{op_ferm} had to be left-handed, but some can be either a 
left-handed doublet or a right-handed quark of a specific flavor. To cover all these scenarios we introduce,
\begin{align}
 c_{u,i} &= \eta_i + (1-\eta_i)\Big(Y_i+ \frac{1}{3} \Big)\,, \nn \\
 c_{d,i} &= \eta_i + (1-\eta_i)\Big(\frac{2}{3} - Y_i \Big)\,.
\end{align}

\subsubsection{One-loop matching}

In the one-loop corrections to the matching, we need to take into account that only the $W$ and $Z$ boson 
(but not the photon) are integrated out, and we need to include the mass-splitting between the $W$ and the $Z$. 
(Since the $W$ and $Z$ mass only enter in the one-loop low scale matching, we can neglect their running.) 
The following relation will be useful,
\begin{align}  \label{eq:couplbr}
  &\al_2 \bT_i \sdt \bT_j + \al_1 \mathbf{Y}_i \mathbf{Y}_j  \\
  &= \al_W (\bT_i^+ \bT_j^- + \bT_i^- \bT_j^+) + \al_Z \bT_i^Z \bT_j^Z + \al_\text{em} \mathbf{Q}_i \mathbf{Q}_j \nn \\
  &= \al_2 (\bT_i \sdt \bT_j - \bT_i^3 \bT_j^3) + (\al_2 + \al_1) \bT_i^Z \bT_j^Z + \al_\text{em} \mathbf{Q}_i \mathbf{Q}_j \,, \nn
\end{align}
where $\bT^\pm = (\bT^1 \pm i \bT^2)/\sqrt{2}$, $\bT^Z = \bT^3 - \sin^2 \theta_W \mathbf{Q}$, $\mathbf{Y}$ is the 
hypercharge and $\mathbf{Q}$ is the charge operator.

In Ref.~\cite{Fuhrer:2010vi}, the low-scale matching was written as a pure quark contribution plus additional 
contributions that involve the Higgs sector. 
We first discuss this quark contribution, which is given in terms of the soft and collinear functions 
of Ref.~\cite{Chiu:2009mg,Chiu:2009ft}, and which we write below as operators acting on quark fields. 
This contribution vanishes for the antisymmetric Higgs operators $O_2^a$ and $O_3$ in \eq{op_higgs}, 
so we can use the same basis as in the tree-level matching, \{$\op_{1A,B}$, $\op_{4A,B,C}$\}.

Using \eq{couplbr}, the soft function for the quarks in the low-scale matching is given 
by~\footnote{We did not separate the cusp and noncusp anomalous dimension in the low-scale matching, 
since we only need the low-scale matching at one-loop order (see table~\ref{tab:counting}).}
\begin{align} \label{eq:D_S}
  D_S&= \frac{\al_W}{2\pi} \log \frac{M_W^2}{\mu^2} \sum_{(ij)}(-\bT_i \sdt \bT_j + \bT_i^3 \bT_j^3) U_{ij} \nn \\
  & \quad + \frac{\al_Z}{2\pi} \log \frac{M_Z^2}{\mu^2} \Big(-\sum_{(ij)} \bT^Z_i \bT^Z_j U_{ij}\Big)
  \,,
\end{align}
where the sum is over pairs $(ij)$ of quarks. 
We can evaluate the $\bT_i \cdot \bT_j$ part of $D_S$ in the unbroken phase of the gauge group,
\begin{align}
  D_S' \equiv -\sum_{(ij)} \bT_i \sdt \bT_j U_{ij}
  = \frac{\ga_S^{SU(2)}}{\Ga(\al_2)} \text{ with } c_{1\pm},c_{2\pm} \to 0\,, \nn \\[-2ex]
\end{align}
and then convert it to the broken basis in \eq{br_basis} using $R^\zero$. 
The remaining terms in \eq{D_S} can only be evaluated in the broken phase, 
so we first multiply by $R^\zero$ to match onto the broken basis. In the broken basis,
\begin{align} \label{eq:D_S3Z}
  D_S^T & \equiv
  -\sum_{(ij)} \bT_i^3 \bT_j^3 U_{ij} \nn \\
  &= \frac{1}{4} \mathrm{diag}\big(c_1\!+\!c_2,c_1\!-\!c_2,c_1\!-\!c_2,c_1\!+\!c_2,
  c_3\!-\!c_1,c_3\!-\!c_1\big)\,, \nn\\
  D_S^Z &\equiv -\sum_{(ij)} \bT_i^Z \bT_j^Z U_{ij} \nn \\
  & = \mathrm{diag}\big(
  g_{u,1}^2 U_{13} + g_{u,1} g_{u,2} c_6 + g_{u,2}^2 U_{24} \nn \\ & \hspace{8.5ex}
  g_{u,1}^2 U_{13} + g_{u,1} g_{d,2} c_6 + g_{d,2}^2 U_{24} \nn \\ & \hspace{8.5ex}
  g_{d,1}^2 U_{13} + g_{d,1} g_{u,2} c_6 + g_{u,2}^2 U_{24} \nn \\ & \hspace{8.5ex}
  g_{d,1}^2 U_{13} + g_{d,1} g_{d,2} c_6 + g_{d,2}^2 U_{24} \nn \\ & \hspace{8.5ex}
  g_u^2 U_{14} + g_u g_d c_7 + g_d^2 U_{23}, \nn \\ & \hspace{8.5ex}
  g_u^2 U_{23} + g_u g_d c_7 + g_d^2 U_{14} \big)\,, 
\end{align}
where $g_{u,i}$ and $g_{d,i}$ denote the coupling of quark $i$ to the $Z$ boson,
\begin{equation}
  g_{u,i} = \frac{1}{2} \eta_i - \frac{2}{3} \sin^2 \theta_W\,, \quad
  g_{d,i} = -\frac{1}{2} \eta_i + \frac{1}{3} \sin^2 \theta_W\,.
\end{equation}
(For $\hO_E$ and $\hO_F$ all the quark fields are left handed, so we suppressed the subscript ``$i$".) 
In \eq{D_S3Z} we used some of the variables $c_i$ from \eq{c_def2} and introduced
\begin{align}
  c_6 &= U_{14} + U_{23} - U_{12} - U_{34}  \,, \nn \\
  c_7 &= U_{13} + U_{24} - U_{12} - U_{34}  \,.
\end{align}

The collinear functions for the quarks in the low-scale matching are
\begin{align}
  D_C^i &= \frac{\al_W}{4\pi} \frac{\eta_i}{2} D(M_W, \bn_i \sdt p_i, \mu) +
  \frac{\al_Z}{4\pi} g_i^2 D(M_Z, \bn_i \sdt p_i, \mu)\,, \nn \\
\end{align}
which is evaluated in the broken basis. Here, $g_i$ is $g_{u,i}$ ($g_{d,i}$) for an up-type (down-type) quark and
\begin{align}
  D(M,\w,\mu) &= 2 \log \frac{\w}{\mu} \log \frac{M^2}{\mu^2} - \frac{1}{2} \log^2 \frac{M^2}{\mu^2} \nn \\ & \quad
  - \frac{3}{2} \log \frac{M^2}{\mu^2} - \frac{5\pi^2}{12} + \frac{9}{4} \,.
\end{align}

The pure quark contribution to the low-scale matching at one loop is thus
\begin{align} \label{eq:D_q}
  D_Q &= v h_n \bigg[\frac{\al_W}{2\pi} \log \frac{M_W^2}{\mu^2} \big(R^\zero \cdot D_S' - D_S^T \cdot R^\zero\big) \\ & \quad
  + \frac{\al_Z}{2\pi} \log \frac{M_Z^2}{\mu^2} D_S^Z \cdot R^\zero 
  + \exp\Big(\sum_i D_C^i\Big) R^\zero \bigg]\,. \nn
\end{align}
In the formulation of Refs.~\cite{Chiu:2009mg,Chiu:2009ft}, the full low-scale matching is exponentiated. 
However, only the collinear functions contain large logarithms, so the soft function does not need to be exponentiated 
to sum the large logarithms. 
This is a convenient simplification because only the soft function contains off-diagonal matrix elements.

We now give explicit expressions for the additional contributions to the low-scale matching derived in 
Ref.~\cite{Fuhrer:2010vi}. For $\op_{1A,B}$, in addition to the quark contribution described above, we get
\begin{equation} \label{eq:D_1} 
 D_1 = vh_n \bigg[\frac 12 \delta R^h + \frac{\Gamma^h}{M_h^2 v}
- \frac{\lambda}{16\pi^2} \sum_j \eta_j^1 B_0(M_j,M_h^2)\bigg]O\,,
\end{equation}
where $O = O_{A,B}$ is the quark part of the operator. 
This contribution is simply an overall factor and requires a factor of $R^\zero$ to convert to the broken basis. 
The finite part of the Higgs wave function $R^h$ and the finite contribution from the tadpole $\Ga^h$ are 
given in \app{RH_GH}. The remaining contribution comes from rescattering, 
where the sum on $j$ runs over $\{h, W, Z\}$ with combinatorial weight $\eta_j^1 = \{3,2,1\}$ and $B_0$ 
is the finite part of the Passarino-Veltman function defined in \eq{PV}.

For $\op_{2A,B,C}$ we do not have a pure quark contribution, due to the antisymmetric combination of scalar fields. 
The only contribution is
\begin{align} \label{eq:D_2}
D_2 &= -\frac{v h_n}{2} \bigg[\frac{\al_W}{4\pi} \log \frac{M_W^2}{\mu^2} \sum_i \Big(U_{ih} + \log \frac{\bn_i \sdt p_i}{\mu}\Big) (\bT_i^a O^a  \nn \\ & \quad
-\! \bT_i^3 O^3) 
 \!+\! \frac{\al_Z}{4\pi} \log \frac{M_Z^2}{\mu^2}\sum_i\! \Big(U_{ih} \!+\! \log \frac{\bn_i \sdt p_i}{\mu}\Big) \bT_i^Z O^3\bigg]\,, 
\end{align}
where $O^a = O^a_{A,B,C}$ is the quark part of the operator. We can evaluate the $\bT_i^a O^a$-piece in the unbroken phase,
\begin{widetext}
\begin{align} \label{eq:D_2p}
  D_2' & \equiv -\sum_i \Big(U_{ih} + \log \frac{\bn_i \sdt p_i}{\mu}\Big) \bT_i^a O^a 
  = 
  \begin{pmatrix}
   c_{2-} + \eta_2 L_{2/4} &  c_{1-} + \eta_1 L_{1/3} & c_{1+} - c_{2+} +\eta_1 \eta_2 L_{13/24} \\
   c_{1-} + \eta_1 L_{1/3}  & c_{2-} + \eta_2 L_{2/4}  & 0
  \end{pmatrix}\,,
\end{align}
\end{widetext}
which is a $2\times3$ matrix in the basis $\{O_{A,B}\}\times\{O^a_{A,B,C}\}$, 
and can be converted to the broken basis using $R^\zero$. The $c_{1\pm}$ and $c_{2\pm}$ were defined in \eq{c_def2}. 
The remaining terms in \eq{D_2} are evaluated in the broken phase. 
First we convert $-O_{A,B,C}^3/2$ to the broken phase by multiplying with $R^\zero$. 
[Note that the factor of $-1/2$ for $O_4^a$ in \eq{Higgs_tree} was absorbed in $R^\zero$.] In the broken basis,
\begin{align} \label{eq:D_3Z}
  D^T & \equiv -\sum_i \Big(U_{ih} + \log \frac{\bn_i \sdt p_i}{\mu}\Big) \bT_i^3 \nn \\
  &= \frac{1}{2}\, \mathrm{diag}\big(
  \eta_1 c_8 + \eta_2 c_9,\, \eta_1 c_8 - \eta_2 c_9,  \nn \\ & \hspace{10.5ex} 
  -\eta_1 c_8 + \eta_2 c_9,\, -\eta_1 c_8 - \eta_2 c_9,  \nn \\ & \hspace{10.5ex} 
  \eta_1 \eta_2 (c_{10}-c_{11}),\, -\eta_1 \eta_2 (c_{10}-c_{11})\big) \nn \\
  D^Z & \equiv -\sum_i \Big(U_{ih} + \log \frac{\bn_i \sdt p_i}{\mu}\Big) \bT_i^Z \nn \\
  &= \mathrm{diag}\big(
  g_{u,1} c_8 + g_{u,2} c_9,\,  g_{u,1} c_8 + g_{d,2} c_9, \nn \\ & \hspace{8.5ex}
  g_{d,1} c_8 + g_{u,2} c_9, \,  g_{d,1} c_8 + g_{d,2} c_9, \nn \\ & \hspace{8.5ex}
  g_u c_{10} + g_d c_{11}, \,  g_d c_{10} + g_u c_{11},
\end{align}
where
\begin{align}
  c_8 &= U_{1h} - U_{3h} + L_{1/3}\,, \quad
  c_9 = U_{2h} - U_{4h} + L_{2/4}\,, \nn \\
  c_{10} &= U_{1h} - U_{4h} + L_{1/4}\,, \quad
  c_{11} = U_{2h} - U_{3h} + L_{2/3}\,.
\end{align}
It is worth pointing out that \eq{D_2} is actually $\mu$-independent, as is clear in \eqs{D_2p}{D_3Z}.

For $\op_{3A,B}$ we do not have a pure quark contribution either. The only contribution is
\begin{equation}
  D_3 = v h_n\,\frac{\al_Z}{4\pi} \log \frac{M_Z^2}{\mu^2} \sum_i \Big(U_{ih} + \log \frac{\bn_i \sdt p_i}{\mu}\Big) \bT_i^Z O\,.
\end{equation}
This can be evaluated by converting the quark part of the operator to the broken phase using $R^\zero$ and 
then applying \eq{D_3Z}. 

For $\op_{4A,B,C}$, in addition to the quark contribution in \eq{D_q}, we get
\begin{align} \label{eq:D_4}
D_4 &= -\frac{vh}{2} \bigg[\bigg(\frac 12 \delta R^h + \frac{\Gamma^h}{M_h^2 v} - \frac{\lambda}{16\pi^2} \sum_j \eta_j^4 B_0(M_j,M_h^2)  \nn \\ & \quad
+  I_h\bigg) O^3
 + \frac{\al_W}{4\pi} \log \frac{M_W^2}{\mu^2}
\sum_i (L_i - U_{ih}) i \eps^{ab3} \bT_i^a O^b \bigg]\,.
\end{align}
The combinatorial weights for the rescattering contribution are now $\eta_j^4 = \{3,-2,1\}$. 
The new contribution $I_h$ is given by
\begin{equation} \label{eq:I_h}
 I_h = \frac{\al_W}{4\pi} \bigg[(2 L_h + 1) \log \frac{M_W^2}{\mu^2} - 1 + f_S\Big(\frac{M_h^2}{M_W^2},1\Big)\bigg]\,,
\end{equation}
where
\begin{align}
 f_S(w,1) &= 1 + \frac{\pi^2}{3} - 2 \sqrt{\frac{w-4}{w}} \tanh^{-1} \sqrt{\frac{w}{w-4}} \\ & \quad
 - 2 \text{Li}_2 \frac{w \!+\!\sqrt{(w\!-\!4)w}}{2} - 2 \text{Li}_2 \frac{w \!-\! \sqrt{(w \!-\! 4)w}}{2}
 \,. \nonumber 
\end{align}
We evaluate the last term in \eq{D_4} in the unbroken phase,
\begin{widetext}
\begin{align} \label{eq:D_eps}
 D^\eps &\equiv \sum_i (L_i - U_{ih}) i \eps^{ab3} \bT_i^a O^b 
 = \begin{pmatrix}
   c_{1+}-\eta_1 L_{13} & 0 & (-c_{2-}+\eta_2 L_{2/4})/2 \\
   0 & c_{2+}-\eta_2 L_{24} & (c_{1-}+\eta_1 L_{3/1})/2 \\
   -c_{2-}+\eta_2 L_{2/4} & c_{1-}+\eta_1 L_{3/1} & (c_{1+}+c_{2+}-\eta_1 L_{13}-\eta_2 L_{24})/2
 \end{pmatrix} \,. 
\end{align}
\end{widetext}
This matrix is given in the basis $\{O^3_{A,B,C}\}\times\{O^a_{A,B,C}\}$. 
We can then convert $-O_{A,B,C}^3/2$ to the broken phase by multiplying with $R^\zero$. 
As for $D_1$, $D_2$ and $D_3$, the explicit $\mu$ dependence cancels out in $D_4$, between \eq{I_h} and \eq{D_eps}. 
These extra pieces therefore do not contain a large logarithm and do not need to be exponentiated 
[see the discussion below \eq{D_q}]. 
We have again used the RPI-III transformation in \eq{RPI} to perform a simple check on these expressions.

\subsection{Electroweak cross section}
\label{subsec:ew}

To obtain the electroweak cross section, we need to square the matrix elements of the operators in \eq{br_basis}, 
sum over quark helicities, flavors, channels and integrate over phase space. Schematically,
\begin{align} \label{eq:si_ew}
\si_\mathrm{EW} &= \frac{1}{2\Ecm^2} \int \frac{\df x_1}{x_1}\, \frac{\df x_2}{x_2}\, \df \Phi_3\, \frac{1}{4} \sum_\mathrm{hel.} \sum_\mathrm{flav.} \sum_\mathrm{chan.} \nn \\ & \quad \times
\sum_{X,Y} f_i(x_1) f_j(x_2)\, \hC_X^* \hC_Y \langle \hO_X \rangle^* \langle \hO_Y \rangle
\,,\end{align}
where $1/(2\Ecm^2)$ is the luminosity factor and $f_i(x)$ denotes the PDF of flavor $i$ evaluated at momentum fraction $x$. 
The three-body phase space $\df \Phi_3$ is  given in \eq{ph_sp}, and an identical particle factor of 1/2 (not shown) 
must be included for symmetric phase-space integrations. 
With ``helicities" we refer to the left- or right-handedness of the fermion fields in the operators, 
which runs over $\{LL,LR,RR\}$ and fixes the spins of the external particles. 
Only $1/2$ of each of the incoming quarks has a given spin, leading to the explicit factor of $1/4$. 
The sum over ``flavors" runs over $\{uu,ud,dd\}$ as well as the generations of up and down-type quarks. 
With ``channels" we mean the assignment of quark flavor and helicity to the lines of the tree-level diagrams 
in Fig.1.

The Wilson coefficients $\hC_X$ are obtained by matching at the high scale, running down to the low scale and 
matching at the low scale,
\begin{equation}
 \hC(\mu_l) = D(\mu_l) \mathcal{U}(\mu_l,\mu_h) C(\mu_h)
 \,,
\end{equation}
where $\mathcal{U}$ solves the RG equation in \eq{RGE}.
In the previous sections we described this in detail for the $t$-channel contribution with incoming quarks, 
and we will extend it to the general case below. 
To discuss the details, some of which are not shown in  \eq{si_ew}, we consider the helicity cases separately.

\subsubsection{Left-Right}

In this case we do not have to worry about identical particles. From the $t$-channel contribution for incoming quarks, 
we can obtain the other channels by a permutation of the momenta. 
For example, if the particles 1 and 2 in the original graph were $u_L$ and $d_R$, we have 
\begin{align} \label{eq:perm_ud}
 u_L\,, d_R:& \quad \id_t\,, (34)_u\,, \nn \\
 u_L\,, \bar u_L:& \quad (23)_s\,, (243)_s\,, \nn \\  
 u_L\,, \bar d_R:& \quad (24)_t\,, (234)_u\,, \nn \\ 
 d_R\,, u_L:& \quad (12)_u\,, (12)(34)_t\,, \nn \\
 d_R\,, \bar u_L:& \quad (132)_u\,, (1432)_t\,, \nn \\
 d_R\,, \bar d_R:& \quad (142)_s\,, (1342)_s\,, \nn \\  
 \bar u_L\,, u_L:& \quad (123)_s\,, (1243)_s\,, \nn \\
 \bar u_L\,, d_R:& \quad (13)_t\,, (143)_u\,, \nn \\
\bar u_L\,, \bar d_R:& \quad (13)(24)_t\,, (1423)_u\,, \nn \\
 \bar d_R\,, u_L: &  \quad (124)_u\,, (1234)_t\,, \nn \\
 \bar d_R\,, d_R:& \quad (14)_s\,, (134)_s\,, \nn \\
\bar d_R\,, \bar u_L:& \quad (14)(23)_t\,, (1324)_u\,.
\end{align}
Here we employ the usual notation for elements of the permutation group $S_4$, e.g.~$(142)$ corresponds 
to $1 \to 4 \to 2 \to 1$, and $\id$ denotes the trivial permutation. 
The subscripts on the permutations indicate whether this contribution corresponds to an $s$, $t$ 
or $u$-channel diagram. 
The permutations are grouped by the type of incoming particle, where we (for simplicity) assumed that 
in the original diagram particle 1 and 2 were an up and down-type quark, respectively. 

Whether the incoming quarks are up or down-type, fixes the flavors of the PDFs except for the generation. 
The electroweak corrections do not depend on the generation and the sum over generations 
differs between the various channels. For example,
\begin{align}\label{eq:LRpdf}
  u_1\,, d_2 \ (t,u): & \ [f_u(x_1) \!+\! f_c(x_1)] [f_d(x_2) \!+\! f_s(x_2) \!+\! f_b(x_2)]
\,.\end{align}

Up to NLL order, we only need the tree-level matrix element of the operators at the low scale. 
For the operators in \eq{br_basis} we find
\begin{equation} \label{eq:op_contr}
  \langle \hO_X \rangle^* \langle \hO_Y \rangle = 4\de_{XY} \times \left\{ 
  \begin{tabular}{ll}
  $s s'$ \quad for & $LL, RR$ \\
  $u_3 u_4$ & $LR, RL$
  \end{tabular} \right.
\,,\end{equation}
which depends on the helicity of the fields for particle 1 and 2, as indicated. 
To obtain the analogue of \eq{op_contr} for any of the permutations, 
we may simply permute the left and right-hand side of this equation. 
Note that due to the nature of the electroweak corrections, $u_L$ and $d_L$ are always calculated simultaneously.

\subsubsection{Right-Right}

For an up-type and a down-type quark, the approach is the same as for LR. 
If the quarks are of the same type, only half the permutation in \eq{perm_ud} remain. For example, 
\begin{align} \label{eq:perm_uu}
 u_R\,, u_R:& \quad \id_t\,, (34)_u\,, \nn \\
 u_R\,, \bar u_R:& \quad (23)_s\,, (234)_u\,, \nn \\  
 & \quad (24)_t\,, (243)_s\,, \nn \\ 
 \bar u_R\,, u_R:& \quad (13)_t\,, (134)_s\,, \nn \\
& \quad (14)_s\,, (143)_u\,, \nn \\
\bar u_R\,, \bar u_R:& \quad (13)(24)_t\,, (1324)_u\,.
\end{align}

For identical quarks and helicities, there are interference contributions. 
Since these are small, we do not include them.

\subsubsection{Left-Left}

For $\hO_A$ and $\hO_D$ in \eq{br_basis}, for which the flavor types are identical, 
the method for LL is the same as for RR. Note that $\hO_C$ is the $(12)(34)$ permutation of $\hO_B$, 
and $\hO_F$ is the $(12)(34)$ permutation of $\hO_E$. We may therefore use the reduced set of permutations 
in \eq{perm_uu} rather than \eq{perm_ud} even though for $\hO_B$, $\hO_C$, $\hO_E$ and $\hO_F$ the flavor types differ. 
(The reason is that the electroweak corrections for $u_L$ and $d_L$ are always calculated simultaneously.)

\subsection{Combining electroweak and QCD corrections}
\label{subsec:QCD}

We will now show that we can multiplicatively include our electroweak corrections in the QCD cross section, 
\begin{equation} \label{eq:si_mult}
\si = \frac{\si_{QCD} \times \si_{EW}}{\si_\mathrm{tree}}\,,
\end{equation}
for any given point in phase space and flavor of the incoming partons. 
Our discussion will be specifically for the VBF process but can be fairly straightforwardly extended to other cases. 
Eq.~\eqref{eq:si_mult} is only true up to NLL order, since the cross section has the following schematic form,
\begin{align} \label{eq:si_logs}
  \log \si &= \sum_n  \underbrace{L(\al_s L)^n + L(\al_w L)^n}_{LL} + \underbrace{(\al_s L)^n + (\al_w L)^n}_{NLL} 
  \nn \\ & \quad
  + \underbrace{\al_s(\al_s L)^n + \al_w(\al_w L)^n + \al_w(\al_s L)^n + \dots}_{NNLL}\,.
\end{align}
where $L$ denotes a large logarithm and $\al_w$ is a weak coupling ($\al_2$ or $\al_1$). 
At NNLL order, the QCD and electroweak corrections get mixed and cannot be calculated separately. 
We will see below explicitly where this breakdown occurs. 
However, one might only care about getting the correct NNLL QCD corrections, 
since the electroweak corrections are smaller.

To discuss how QCD and electroweak corrections can be combined, 
we will assume that the QCD corrections are also calculated using SCET. 
The QCD corrections only affect the quarks, for which the possible color structures are given by
\begin{equation}
 \bar \Psi_3 \ga^\mu T^a_{SU(3)} \Psi_1 \bar \Psi_4 \ga_\mu T^a_{SU(3)} \Psi_2\,, \quad
 \bar \Psi_3 \ga^\mu \Psi_1 \bar \Psi_4 \ga_\mu \Psi_2\,.
\end{equation}
In this equation we suppressed the $SU(2)$ structure, since the $SU(3)$ and $SU(2)$ generators act in different spaces. 
Including these color structures doubles the basis in \eq{op_ferm}, 
which we will view as a tensor product of the $SU(2)$ and $SU(3)$ structures. 

At NLL order, the high-scale matching is performed at tree-level. 
Since the diagrams are purely electroweak, it trivially factors,
\begin{equation} \label{eq:C_factor}
C(\mu_h) = C_{SU(2)}(\mu_h) \otimes C_{SU(3)}(\mu_h)
\,.
\end{equation}
The one-loop QCD corrections to the high-scale matching differ between the $s$, $t$ and $u$-channel diagrams. 
Thus \eq{C_factor} no longer holds when different channels are combined, which is inevitable due to interferences 
between channels. 

Up to NLL order, the anomalous dimension completely separates
\begin{equation} \label{eq:ga_ew_QCD}
  \ga = \ga_{SU(2)} \otimes \id_{SU(3)} + \id_{SU(2)} \otimes \ga_{SU(3)}
  \,.
\end{equation}
Here $\ga_{SU(2)}$ and $\ga_{SU(3)}$ denote the usual electroweak and QCD anomalous dimensions, 
i.e.~$\ga_{SU(2)}$ only acts on the $SU(2)$ structures and only involves $\al_2, \al_1$. 
At one-loop order \eq{ga_ew_QCD} is trivial, because the loop is either QCD or electroweak. At two-loop order, 
\eq{ga_ew_QCD} no longer holds in general, but it does still hold for the cusp anomalous dimension. 
Since gluons and electroweak gauge bosons only couple through fermions and act in different spaces, 
a mixed contribution would require a $C_F^2$, with one $C_F$ from $SU(3)$ and another from $SU(2)$. 
However, the two-loop cusp does not contain a $C_F^2$ term. At three-loop order, diagrams such as Fig.2 arise, 
that produce a $C_F^2$ in the cusp anomalous dimension and violate the form in \eq{ga_ew_QCD}. 
Thus \eq{ga_ew_QCD} holds up to NLL order but not beyond.
Integrating \eq{ga_ew_QCD} from the high scale to the low scale leads to an evolution factor
\begin{equation} \label{eq:U_factor}
 \mathcal{U}(\mu_h,\mu_l) =  \mathcal{U}_{SU(2)}(\mu_h,\mu_l) \otimes \mathcal{U}_{SU(3)}(\mu_h,\mu_l)\,,
\end{equation}
which is just the tensor product of the electroweak and QCD evolution.

\begin{figure}[t] \label{fig:cusp}
\centering
\includegraphics[width=0.22\textwidth]{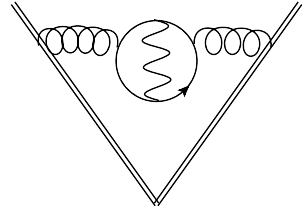}
\caption{Three-loop contribution to the cusp anomalous dimension that violates \eq{ga_ew_QCD}. The double lines denote Wilson lines.}
\end{figure}

The low-scale matching only affects the electroweak basis,
\begin{equation} \label{eq:D_factor}
  D(\mu_l) = D_{SU(2)}(\mu_l) \otimes \id_{SU(3)}
  \,.
\end{equation}
At one-loop order, $D_{SU(2)}$ is trivially the same as in the pure electroweak case. 
For the large logarithm in the low-scale matching it also holds at two-loop order, 
because its coefficient is the cusp anomalous dimension. 
Since the low-scale matching is needed at one lower order than the running (see table~\ref{tab:counting}), 
the low scale matching even factors at NNLL.

In \subsec{ew}, we took the tree-level matrix element of the operators at the low scales, 
which is sufficient for the pure electroweak corrections up to NLL order. 
The situation is more complicated for QCD, which becomes strongly coupled at low scales. 
QCD can also have its own series of large double logarithms that depend on the measurement on the hadronic final state. 
An example of a measurement for an $N$-jet signal at hadron colliders is the 
event shape $N$-jettiness $\tau_N$~\cite{Stewart:2010tn}. 
Requiring $\tau_N \leq \tau_N^\mathrm{cut} \ll 1$ defines an exclusive $N$-jet measurement and 
induces large QCD logarithms $\al_s^n \log^m \tau_N$ (with $m\leq 2n$) from vetoing additional jets. 
In the presence of QCD corrections we still square the operators and get the tree-level weights 
in \eq{op_contr}. Since there is no dependence on the quark flavor in the perturbative QCD corrections, 
these directly factor. 
The QCD corrections do depend on the quark flavor through the PDFs, so \eq{si_mult} only holds when the 
flavor of the incoming quarks is fixed. 
We point out that at one-loop order there are corrections involving the gluon PDF, 
but this contribution is equal for all quark flavors and thus preserves the factorization of 
QCD and electroweak corrections.

\section{Numerical results}
\label{sec:results}

In this section we report our numerical results for the integrated cross section.
The couplings and parameters of the standard model have been set at the electroweak scale $\mu=M_Z$ according
to the data of Ref.\cite{PDG}. Two-loop beta functions are used\cite{Arason} 
for their running up to the high energy scale $\mu_h$. The Higgs mass is assumed to be $M_H=125$ GeV.

The low energy matching scale is chosen to be $\mu_l=M_Z$. The cross section should not depend on it,
since $\mu_l$ is not a physical scale. We checked that the dependence is very small because of 
a cancellation between the running and the low-scale matching. However, the electromagnetic part of the weak 
interactions does not get integrated out at the low scale, and the cancellation can never be perfect.
In fact the scale dependence is larger in the Left-Right and Right-Right processes where the electromagnetic corrections 
become more important. That is not a real issue since the cross section is dominated by the Left-Left terms,
and the low scale dependence of the total cross section is negligible. With our choice $\mu_l=M_Z$,  
in the low-scale matching only the collinear function is needed, the remainder being less than 1\%. 
That is because the remaining contribution largely cancel.  We keep all the small terms  in our numerical calculation
anyway.
Away from $M_Z$, these terms become important since they cancel the low scale dependence of the
running. 

As shown in table~\ref{tab:counting}, at NLL order only the tree level high energy matching is required, and the one-loop
matching terms can be neglected. As for any tree-level calculation, the result does depend on the high energy scale
$\mu_h$ because there is no cancellation between the running and the high energy matching at tree-level.
Thus, as usual, the high scale must be thoroughly chosen in order to keep the neglected one-loop terms small enough.
Typically, the one-loop matching terms contain the logarithmic terms $\ln \hs_{ij}/\mu_h^2$, thus whenever
the $\hs_{ij}$ are of the same size, we take $\mu_h^2\approx \hs_{ij}$ and the one-loop matching terms can be safely
neglected, as they do not contain large logarithms. In VBF we are mainly interested in the $t$ channel,  and the
relevant logarithms that appear in the one-loop matching are $\ln ({-t_3}/\mu_h^2)$ and $\ln ({-t_4}/\mu_h^2)$.
However there are small phase space regions where $t_3\ll t_4$ or $t_4\ll t_3$, and the choice of the correct
high scale turns out to be ambiguous. The best way out would be a full calculation of the one-loop high energy
matching terms, but that would be beyond the aim of the present paper. Moreover, avoiding difficult electroweak
loop calculations was one of the motivations of the present study. It is worth mentioning that for VBF
the same ambiguous choice of the high scale is also encountered in the tree level calculation of the cross
section whenever $t_3$ and $t_4$ are not of the same size.

\begin{figure}[t] \label{fig:relall}
\centering
\includegraphics[width=0.37\textwidth,angle=-90]{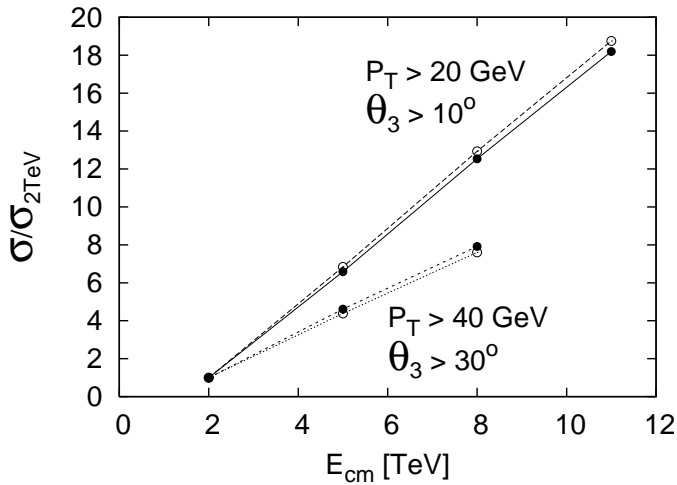}
\caption{Tree-level cross section according to \eq{si_ew} but with no radiative corrections included,
divided by the result for $\Ecm=2$ TeV, as a function of the center-of-mass energy $\Ecm$ (filled circles).
The output of HAWK is reported for comparison (open circles). Two sets of cuts have been used:
$p_T>20$ GeV, $\theta_3> 10^o$, $\theta_4< 170^o$ (upper data); $p_T>40$ GeV, $\theta_3> 30^o$, $\theta_4< 150^o$
(lower data).}
\end{figure}

As a good interpolation between phase space regions where $t_3\approx t_4$ and regions where one of the variables
is small, we set the high scale to the minimum between $M_Z$ and the geometric average $\mu_h^2=\sqrt{t_3 t_4}$.
At this scale the sum of logarithmic terms $\ln ({-t_3}/\mu^2)+\ln ({-t_4}/\mu^2)$ reaches its minimum in the one-loop
matching. 
Moreover this choice has the merit of stopping the running whenever one of the Mandelstam variables is too small.
We will discuss the sensitivity of the result to the choice of the high scale, and show that it is comparable
to the sensitivity of the standard tree-level cross section.~\footnote{
We would like to stress once more that, at variance with the variational approaches 
to the electroweak sector\cite{Siringo:2008,Siringo:2010,Siringo:2012}, 
the principle of minimum sensitivity here does
not apply even if $\mu_h$ is not a physical scale. In fact the sensitivity only measures the dependence
on $\mu_h$ of the neglected terms, and this dependence might be high even when the neglected terms are small.
The best choice of $\mu_h$ should make the omitted terms small, as for the method of minimal variance\cite{Siringo:2005}.}

The aim of the paper is a test of the method by a comparison with the one-loop fixed order perturbative
calculation of the code HAWK that is based on the work of Ref.\cite{Ciccolini:2007jr,Ciccolini:2007ec}. 
In our comparison we adopt larger cuts than phenomenologically required, in order to be
sure that we are on safe grounds for the SCET approximation.
We use CTEQ6 PDFs\cite{PDF} and neglect the very small contribution of t and b quarks. Moreover in both codes
we neglect the s-channel contribution and interference terms that are known to be small with VBF cuts.

\begin{figure}[t] \label{fig:terms}
\centering
\includegraphics[width=0.37\textwidth,angle=-90]{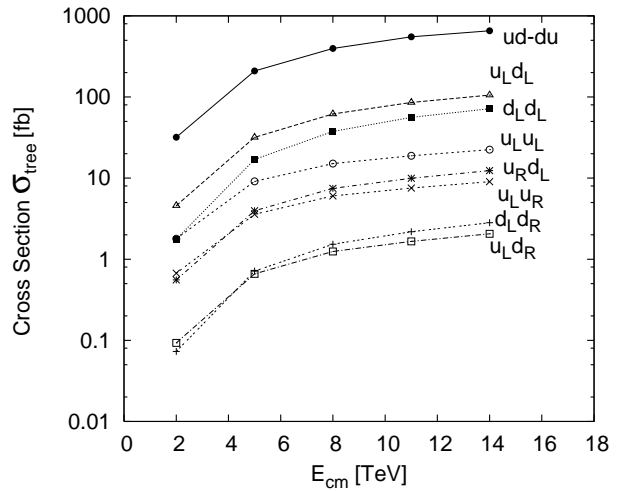}
\caption{Single terms contibuting to the tree-level cross section as a function of the center-of-mass energy $\Ecm$.
Cuts are: $p_T>20$ GeV, $\theta_3> 10^o$, $\theta_4< 170^o$.}
\end{figure}

At tree level we find a fair agreement between the codes, provided that the masses of the vector gauge bosons are
restored in the denominators of the coefficients in \eq{c_def}. While that is not so relevant when relative
cross sections are reported with large cuts, the effect of masses becomes crucial when the total cross section
is evaluated by integrating over the whole phase space. In relative terms, the tree-level cross section, without
any radiative correction, is reported in Fig.3 for several values of the center-of-mass energy $\Ecm$ and
for different cuts, as a test of the integration over the PDFs according to \eq{si_ew}. 

Hereafter we adopt the following choice of cuts on angles 
and transverse moment: $\theta_3> 10^o$, $\theta_4< 170^o$, $p_T>20$ GeV.
The single terms contributing to the cross section at tree-level are reported in Fig.4. The $ud\to du$
process is the dominating one, followed by the other Left-Left terms. The Left-Right terms are two order of
magnitude smaller, while the Right-Right terms are negligible and have not been included in the calculation.
A sum over generations of the up- and down-type quarks is included through the PDFs in the terms of Fig.4.

\begin{figure}[t] \label{fig:ewcorr}
\centering
\includegraphics[width=0.32\textwidth,angle=-90]{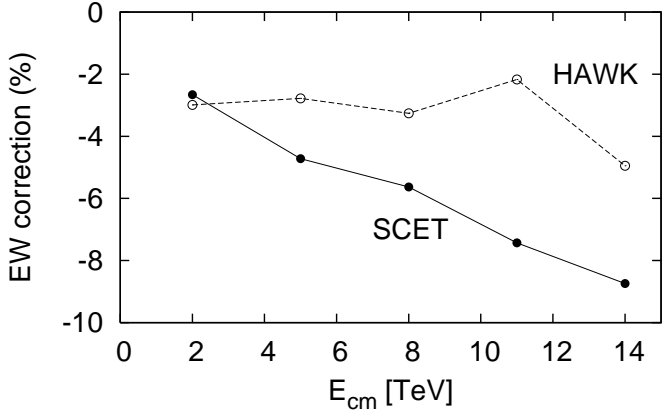}
\caption{Relative electroweak correction $[\sigma_{EW}-\sigma_{TREE}]/\sigma_{tree}$
as a function of the center-of-mass energy $\Ecm$ evaluated by \eq{si_ew} in
this paper (filled circles), compared with the output of HAWK (open circles).
Cuts are: $p_T>20$ GeV, $\theta_3> 10^o$, $\theta_4< 170^o$.}
\end{figure}

\begin{figure}[t] \label{fig:theta}
\centering
\includegraphics[width=0.32\textwidth,angle=-90]{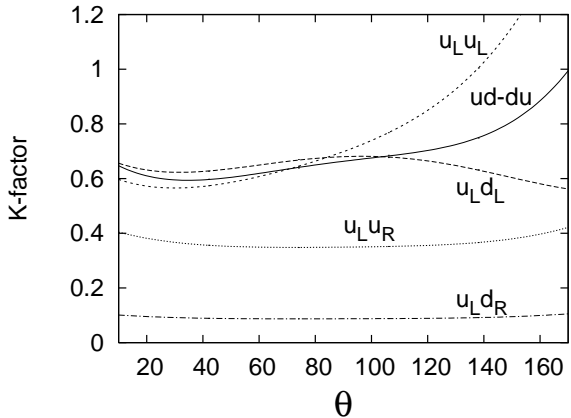}
\caption{The K-factor $K=\sigma_{EW}/\sigma_{tree}$ as a function of $\theta=\theta_3=180^o-\theta_4$,
at $\varphi=90^o$, $x_1=x_2=0.5$, $\Ecm=14$ TeV and $E_3=x_1 x_2\Ecm/2$.}
\end{figure}

The electroweak corrections have been evaluated integrating in \eq{si_ew} with the same cuts. 
In order to compare with HAWK no QCD corrections have been included in both codes, and only
virtual electroweak correction are considered.
We find large electroweak corrections in wide sectors of the phase space, 
expecially for the Left-Right processes that get suppressed by
even more than 90\% at some spots (see Figs. 6 and 7). 
However, after integrating with PDFs, the corrections are not dramatic.
In fact the corrections are smaller for the $ud\to du$ process which dominates in the cross section,
while the highly suppressed terms, like $u_Ld_R$, have a very small weight on the total result.

\begin{figure}[t] \label{fig:phi}
\centering
\includegraphics[width=0.32\textwidth,angle=-90]{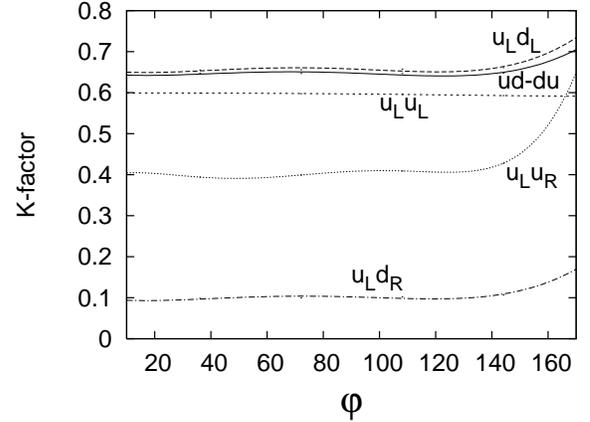}
\caption{The K-factor $K=\sigma_{EW}/\sigma_{tree}$ as a function of $\varphi$ at $\theta_3=10^o$, 
$\theta_4=170^o$, $x_1=x_2=0.5$, $\Ecm=14$ TeV and $E_3=x_1 x_2\Ecm/2$.}
\end{figure}

In Fig.5 we show the relative electroweak correction, defined as the ratio 
\begin{equation} \label{eq:relsig}
\frac{\Delta\sigma}{\sigma}=
\frac{\sigma_{EW}-\sigma_{tree}}{\sigma_{tree}}
\,,
\end{equation}
as a function of the center-of-mass energy.
The correction is less than 3\% at $\Ecm=2$ TeV, in good agreement with the output of HAWK.
At larger energies the correction grows faster than predicted by HAWK, reaching 9\% at the
LHC energy $\Ecm=14$ TeV, to be compared with 5\% predicted by HAWK.
At the higher energies, the larger contribution to the electroweak correction comes
from the running of the coefficients by the anomalous dimension. Thus we argue that fixed-order
perturbative calculations might miss part of the correction at the LHC energy scale.

It is instructive  to see how the single terms behave in the phase space, and for each of them
we define a sort of K-factor as 
\begin{equation} \label{eq:kfac}
K=
\frac{\sigma_{EW}(\theta_3,\theta_4,\varphi)}{\sigma_{tree}(\theta_3,\theta_4,\varphi)}
\,,
\end{equation}
where the differential cross sections are evaluated for fixed values of $x_1$,$x_2$, $E_3$ and $\Ecm$,
and for a given set of angles. In the physically relevant range of the parameters we do not find
any important dependence on the angles, and the K-factor is almost flat at small angles.
For instance in Fig.6 we show the dependence on $\theta_3$, $\theta_4$. The K-factor is reported
as a function of $\theta=\theta_3$ with $\theta_4=\pi-\theta_3$ and $\varphi=\pi/4$ at a typical
set of parameters: $\Ecm=14$ TeV, $x_1=x_2=0.5$ and $E_3=x_1x_2\Ecm/2$. For $\theta<90^o$ the dependence
on $\theta$ is very small and is not expected to have any important relevance on the angular distribution
of the jets. It is remarkable the very large suppression of the $u_L d_R$ process which has no
relevant effect on the total cross section that is dominated by the Left-Left processes.
The dependence on the azimuthal angle $\varphi$ seems to be even smaller, but becomes more relevant
when $\varphi$ approaches $180^o$, indicating that at large values of $\varphi\approx 180^o$  the cross section
gets enhanced by the electroweak corrections, as predicted by the NLO calculation of Ref.\cite{Ciccolini:2007ec}. 
For instance in Fig.7 the dependence
on $\varphi$ is shown at small angles $\theta_3=10^o$, $\theta_4=170^o$ and for the same set of parameters
as before. 

\begin{figure}[t] \label{fig:test}
\centering
\includegraphics[width=0.40\textwidth,angle=-90]{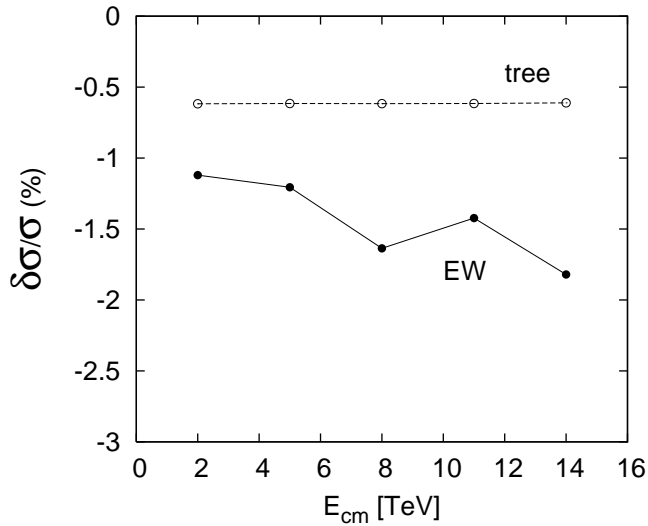}
\caption{The relative change of the cross section in the process $ud\to du$ is reported for
a 10\% increase of the high scale $\mu_h$ (filled circles). 
The tree-level result is shown for comparison (open circles).
Cuts are the same as in Fig. 5.}
\end{figure}

As discussed above, our result must depend on the high scale $\mu_h$ because the tree-level
matching cannot cancel the dependence. The same problem arises in the calculation of the
tree-level cross section, and one takes for granted that the choice of the high scale must
keep the logarithms small in the higher order terms of the perturbative expansion.
However, even if our choice for the high scale seems to work well, 
it would
be desirable to see how the results depend on the choice of $\mu_h$,
expecially at low energies
where the Mandelstam variables might be comparable with the low scale $\mu_l=M_z$.
We can check this dependence
on the most dominant term in the cross section, namely the $ud\to du$ process, and we can compare
with the dependence of the tree-level cross section.

In Fig.8 the relative change $\delta \sigma/ \sigma$ is reported for a ten per cent increase of the
high scale $\delta \mu_h/\mu_h=0.1$. The change of the tree-level cross section is reported for
comparison, and is evaluated by taking  $\mu_h$ as renormalization scale. The tree-level cross section
depends on the high scale because of the running of the couplings, and we find a steady -0.6\% change
for all the energies. As shown in Fig.8, when the electroweak corrections are included the cross section
dependence increases with the increase of energy, going from -1.2\% to -1.8\%.
This dependence is also negative, and remains comparable in size to the tree-level value.
Moreover these dependences cancel each other in the relative electroweak correction $\Delta\sigma/\sigma$
which we reported in Fig.5. In fact, differenziating \eq{relsig} and assuming $\sigma_{tree}\approx\sigma_{EW}$
\begin{equation} \label{eq:delta}
\delta\left(\frac{\Delta \sigma}{\sigma}\right)\approx
\left[\frac{\delta \sigma_{EW}}{\sigma_{EW}}-\frac{\delta \sigma_{tree}}{\sigma_{tree}}
\right]
\,.
\end{equation}
Thus the overall effect on the relative correction goes from -0.6\% to -1.2\% which seems to be more than acceptable
if compared with the natural -0.6\% change of the tree-level cross section.

\section{Conclusions}
\label{sec:concl}

Even if VBF is the second largest production channel for the Higgs boson, it is
a clean and pure electroweak process, and its precise
measurement is very important for constraining the Higgs couplings and thus 
identifying the nature of the Higgs sector. A full control of the radiative
corrections might be also relevant for ruling out extensions of the standard model
or exotic processes that have been predicted in the Higgs sector\cite{Siringo:2000,Siringo:2002}

We have shown that SCET and the method of Refs.~\cite{Chiu:2009mg,Chiu:2009ft} may be used 
to directly obtain electroweak corrections 
for generic processes without the need to perform 
difficult electroweak loop calculations that would have to be carried out for each individual process.
The electroweak corrections have been obtained in analytical form, and can be easily inserted in the software packages 
that have been developed for computing QCD corrections to cross sections. 
This is one of the first tests of the formalism 
at the level of a hadronic cross section, and demonstrates the viability of the method that
has the merit of providing a resummation of the large dominating logarithms at
LHC energies. 
Moreover, in the special case of VBF, the amplitude is explicitly proportional to the VEV, 
so the effective field theory operator is not a gauge singlet and standard resummation methods do not apply. 
Thus VBF provides a very interesting test of the method, besides the phenomenological relevance of the
process in the study of the Higgs boson at LHC.

In our test we used the extension to the VBF process as derived in Ref.~\cite{Fuhrer:2010vi}, 
and compared the integrated cross section with the NLO output of the code HAWK, based on
the calculation of Ref.\cite{Ciccolini:2007jr,Ciccolini:2007ec}.

We find large electroweak corrections in wide sectors of the phase space, 
expecially for the Left-Right processes where the corrections reach 90\%.
They mostly arise from the RG running of the Wilson coefficients of the
effective operators.
However, after integrating with PDFs, the corrections are not dramatic.
In fact the corrections are smaller for the $ud\to du$ process which dominates in the cross section,
while the highly suppressed terms, like $u_Ld_R$, have a very small weight on the total result.

Below 2 TeV our results are in fair agreement with HAWK, while at higher energies the SCET formalism
predicts corrections that are slightly larger, and grow up with increasing the center-of-mass
energy. Since the larger effect comes
from the running, we argue that fixed-order
perturbative calculations might miss part of the correction at the LHC high energy scale.
We also explored the angular dependence of the electroweak correction which is small and in
general agreement with NLO calculations\cite{Ciccolini:2007jr,Ciccolini:2007ec}.

While we checked that the sensitivity to the high scale $\mu_h$ is not too high, and is comparable to
the sensitivity of the tree-level cross section, we are aware that the corrections must depend on the choice 
of $\mu_h$ because no cancellation is provided by the tree-level high scale matching, and that
a thoroughly choice for $\mu_h$ is important. Our choice of a geometric average $\mu^2_h=\sqrt{t_3t_4}$
seems to work well at low energies, and is expected to work even better at higher energies since the 
Mandelstam variables get much larger than the electrowek scale $M_Z$. A full one-loop calculation of the
high energy matching would be required in order to test the choice, but that goes out of the aim of the present paper
and will require further work.

\acknowledgments

We thank A.~Manohar for helpful discussions and feedback on the manuscript.
A special acknowledgement goes to W.J.~Waalewijn for sharing with us his unpublished
results on VBF, for his feedback and suggestions on the manuscript, and for having
brought to our attention the argument on factorization of QCD and electroweak 
corrections as reported in \subsec{QCD}.

\appendix

\section{Passarino-Veltman functions}
\label{app:PV}

The Passarino-Veltman equal-mass functions\cite{Passarino:1978jh} $A_0$, $B_0$ and $AB_0$
are given by the following analytical integral representations (with the normalization of Ref. \cite{Fleischer:1980ub})
\begin{align} \label{eq:PV}
\frac{-i}{16\pi^2}A_0(m)&=\int\!\! \frac{\df^dk}{(2\pi)^d} \!\frac{1}{k^2-m^2+i0} 
\,,\nn \\
\frac{i}{16\pi^2}B_0(m,p^2)
&=\int\!\! \frac{\df^dk}{(2\pi)^d}
\frac{1}{(k^2\!-\!m^2\!+\!i0)[(k\!+\!p)^2\!-\!m^2\!+\!i0]} 
\,, \nn \\
AB_0(m,s) &= \frac{\partial}{\partial s} B_0(m,s)
\,.
\end{align}

In the $\overline{\mathrm{MS}}$ scheme, the finite part of these functions are given by
\begin{align}
A_0(m) &=-m^2\bigg(1-\log{\frac{m^2}{\mu^2}}\bigg) 
\,, \nn \\
B_0(m,s) &=2-\log{\frac{m^2}{\mu^2}}
-2\sqrt{4z-1}\tan^{-1}{\frac{1}{\sqrt{4z-1}}} 
\,, \nn \\
AB_0(m,s) &=\frac{1}{s}\bigg(
\frac{4z}{\sqrt{4z-1}}\tan^{-1}\frac{1}{\sqrt{4z-1}}-1\bigg)
\,,
\end{align}
where $z=m^2/s$. These functions are real if $s<4m^2$, while for $s>4m^2$ they acquire an imaginary part and can be explicitly written as
\begin{align}
B_0(m,s)&=2\!-\!\log{\frac{m^2}{\mu^2}}
\!-\!\sqrt{1\!-\!4z}\left[
\log\frac{1\!+\!\sqrt{1\!-\!4z}}{1\!-\!\sqrt{1\!-\!4z}}
\!-\!i\pi\right]
\,,\nn \\
AB_0(m,s)&=\frac{1}{s}\left[
\frac{2z}{\sqrt{1\!-\!4z}}
\log\frac{1\!-\!\sqrt{1\!-\!4z}}{1\!+\!\sqrt{1\!-\!4z}}\!-\!1\right]
\!+\!\frac{2\pi i\, z}{s\sqrt{1-4z}}\,.
\end{align}

\section{Explicit form of Higgs wave function $R^h$ and tadpole $\Ga^h$}
\label{app:RH_GH}

The finite part of the Higgs wave function $R^h$ and the scalar tadpole graph $\Gamma^h$ appear in the low-scale matching in \eqs{D_1}{D_4}. For completeness we give their explicit expressions here, which can for instance be found in Ref.~\cite{Fleischer:1980ub}. 

In $R_{\xi=1}$ gauge, the Higgs wave function $\de R^h = R^h - 1$ is given by 
\begin{align}
\de R^h &=\frac{1}{16\pi^2 v^2}\Big[2M_Z^2B_0(M_Z,M_h^2)+
4M_W^2B_0(M_W,M_h^2)
\nn \\ & \quad
-\frac{9}{2}M_h^4 AB_0(M_h,M_h^2) - 2m_t^2B_0(m_t,M_h^2) 
\nn \\ & \quad 
+ \Big(2M_h^2M_Z^2 - \frac{1}{2}M_h^4 - 6M_Z^4\Big)
AB_0(M_Z,M_h^2)
\nn \\ & \quad 
+ (4M_h^2M_W^2-M_h^4-12M_W^4)
AB_0(M_W,M_h^2) 
\nn \\ & \quad
-(2M_h^2m_t^2-8m_t^4)AB_0(m_t,M_h^2)\Big]
\,.
\end{align}
Here, $m_t$ is the mass of the top quark, and the contribution from the other fermions have been neglected. The Passarino-Veltman functions $B_0$ and $AB_0$ are given in \app{PV}. Note that while only the real part of the wave function can enter in the renormalization, the full complex value must be kept in the low-scale matching.

The tadpole contribution $\Gamma^h$ is related to the shift $\delta v$ of the VEV. By a self-consistent evaluation it can be written in $R_{\xi=1}$ gauge as
\begin{align}
\frac{\Gamma^h}{M_h^2 v} = \frac{\delta v}{v} 
&= \frac{1}{16\pi^2 v^2 M_h^2}\bigg[
-2M_z^4-4M_W^4
\\ & \quad
-\frac{3}{2}M_h^2A_0(M_h)
-(3M_Z^2+\frac{1}{2}M_h^2)A_0(M_Z)
\nn \\ & \quad
-(6M_W^2+M_h^2)A_0(M_W) + 4m_t^2A_0(m_t) \bigg]
\,, \nn
\end{align}
where again only the fermionic term for the top quark has been retained.

\bibliography{VBF}

\begin{thebibliography}{46}
\expandafter\ifx\csname natexlab\endcsname\relax\def\natexlab#1{#1}\fi
\expandafter\ifx\csname bibnamefont\endcsname\relax
  \def\bibnamefont#1{#1}\fi
\expandafter\ifx\csname bibfnamefont\endcsname\relax
  \def\bibfnamefont#1{#1}\fi
\expandafter\ifx\csname citenamefont\endcsname\relax
  \def\citenamefont#1{#1}\fi
\expandafter\ifx\csname url\endcsname\relax
  \def\url#1{\texttt{#1}}\fi
\expandafter\ifx\csname urlprefix\endcsname\relax\def\urlprefix{URL }\fi
\providecommand{\bibinfo}[2]{#2}
\providecommand{\eprint}[2][]{\url{#2}}

\bibitem[{\citenamefont{Duhrssen et~al.}(2004)\citenamefont{Duhrssen,
  Heinemeyer, Logan, Rainwater, Weiglein et~al.}}]{Duhrssen:2004cv}
\bibinfo{author}{\bibfnamefont{M.}~\bibnamefont{Duhrssen}},
  \bibinfo{author}{\bibfnamefont{S.}~\bibnamefont{Heinemeyer}},
  \bibinfo{author}{\bibfnamefont{H.}~\bibnamefont{Logan}},
  \bibinfo{author}{\bibfnamefont{D.}~\bibnamefont{Rainwater}},
  \bibinfo{author}{\bibfnamefont{G.}~\bibnamefont{Weiglein}},
  \bibnamefont{et~al.}, \bibinfo{journal}{Phys.Rev.}
  \textbf{\bibinfo{volume}{D70}}, \bibinfo{pages}{113009}
  (\bibinfo{year}{2004}), \eprint{hep-ph/0406323}.

\bibitem[{\citenamefont{Zeppenfeld et~al.}(2000)\citenamefont{Zeppenfeld,
  Kinnunen, Nikitenko, and Richter-Was}}]{Zeppenfeld:2000td}
\bibinfo{author}{\bibfnamefont{D.}~\bibnamefont{Zeppenfeld}},
  \bibinfo{author}{\bibfnamefont{R.}~\bibnamefont{Kinnunen}},
  \bibinfo{author}{\bibfnamefont{A.}~\bibnamefont{Nikitenko}},
  \bibnamefont{and}
  \bibinfo{author}{\bibfnamefont{E.}~\bibnamefont{Richter-Was}},
  \bibinfo{journal}{Phys.Rev.} \textbf{\bibinfo{volume}{D62}},
  \bibinfo{pages}{013009} (\bibinfo{year}{2000}), \eprint{hep-ph/0002036}.

\bibitem[{\citenamefont{Bauer et~al.}(2000)\citenamefont{Bauer, Fleming, and
  Luke}}]{BFL}
\bibinfo{author}{\bibfnamefont{C.~W.} \bibnamefont{Bauer}},
  \bibinfo{author}{\bibfnamefont{S.}~\bibnamefont{Fleming}}, \bibnamefont{and}
  \bibinfo{author}{\bibfnamefont{M.~E.} \bibnamefont{Luke}},
  \bibinfo{journal}{Phys. Rev.} \textbf{\bibinfo{volume}{D63}},
  \bibinfo{pages}{014006} (\bibinfo{year}{2000}), \eprint{hep-ph/0005275}.

\bibitem[{\citenamefont{Bauer et~al.}(2001)\citenamefont{Bauer, Fleming,
  Pirjol, and Stewart}}]{SCET1}
\bibinfo{author}{\bibfnamefont{C.~W.} \bibnamefont{Bauer}},
  \bibinfo{author}{\bibfnamefont{S.}~\bibnamefont{Fleming}},
  \bibinfo{author}{\bibfnamefont{D.}~\bibnamefont{Pirjol}}, \bibnamefont{and}
  \bibinfo{author}{\bibfnamefont{I.~W.} \bibnamefont{Stewart}},
  \bibinfo{journal}{Phys. Rev.} \textbf{\bibinfo{volume}{D63}},
  \bibinfo{pages}{114020} (\bibinfo{year}{2001}), \eprint{hep-ph/0011336}.

\bibitem[{\citenamefont{Bauer and Stewart}(2001)}]{SCET2}
\bibinfo{author}{\bibfnamefont{C.~W.} \bibnamefont{Bauer}} \bibnamefont{and}
  \bibinfo{author}{\bibfnamefont{I.~W.} \bibnamefont{Stewart}},
  \bibinfo{journal}{Phys. Lett.} \textbf{\bibinfo{volume}{B516}},
  \bibinfo{pages}{134} (\bibinfo{year}{2001}), \eprint{hep-ph/0107001}.

\bibitem[{\citenamefont{Bauer et~al.}(2002)\citenamefont{Bauer, Pirjol, and
  Stewart}}]{BPS}
\bibinfo{author}{\bibfnamefont{C.~W.} \bibnamefont{Bauer}},
  \bibinfo{author}{\bibfnamefont{D.}~\bibnamefont{Pirjol}}, \bibnamefont{and}
  \bibinfo{author}{\bibfnamefont{I.~W.} \bibnamefont{Stewart}},
  \bibinfo{journal}{Phys. Rev.} \textbf{\bibinfo{volume}{D65}},
  \bibinfo{pages}{054022} (\bibinfo{year}{2002}), \eprint{hep-ph/0109045}.

\bibitem[{\citenamefont{Chiu et~al.}(2008)\citenamefont{Chiu, Golf, Kelley, and
  Manohar}}]{Chiu:2007dg}
\bibinfo{author}{\bibfnamefont{J.-y.} \bibnamefont{Chiu}},
  \bibinfo{author}{\bibfnamefont{F.}~\bibnamefont{Golf}},
  \bibinfo{author}{\bibfnamefont{R.}~\bibnamefont{Kelley}}, \bibnamefont{and}
  \bibinfo{author}{\bibfnamefont{A.~V.} \bibnamefont{Manohar}},
  \bibinfo{journal}{Phys. Rev.} \textbf{\bibinfo{volume}{D77}},
  \bibinfo{pages}{053004} (\bibinfo{year}{2008}), \eprint{arXiv:0712.0396}.

\bibitem[{\citenamefont{Chiu et~al.}(2009)\citenamefont{Chiu, Fuhrer, Kelley,
  and Manohar}}]{Chiu:2009mg}
\bibinfo{author}{\bibfnamefont{J.-y.} \bibnamefont{Chiu}},
  \bibinfo{author}{\bibfnamefont{A.}~\bibnamefont{Fuhrer}},
  \bibinfo{author}{\bibfnamefont{R.}~\bibnamefont{Kelley}}, \bibnamefont{and}
  \bibinfo{author}{\bibfnamefont{A.~V.} \bibnamefont{Manohar}},
  \bibinfo{journal}{Phys. Rev.} \textbf{\bibinfo{volume}{D80}},
  \bibinfo{pages}{094013} (\bibinfo{year}{2009}), \eprint{arXiv:0909.0012}.

\bibitem[{\citenamefont{Denner and Pozzorini}(2001)}]{Pozzorini1}
\bibinfo{author}{\bibfnamefont{A.}~\bibnamefont{Denner}} \bibnamefont{and}
  \bibinfo{author}{\bibfnamefont{S.}~\bibnamefont{Pozzorini}},
  \bibinfo{journal}{Eur. Phys. J.} \textbf{\bibinfo{volume}{C18}},
  \bibinfo{pages}{461} (\bibinfo{year}{2001}), \eprint{arXiv:hep-ph/0010201v3}.

\bibitem[{\citenamefont{Accomando et~al.}(2002)\citenamefont{Accomando, Denner,
  and Pozzorini}}]{Pozzorini2}
\bibinfo{author}{\bibfnamefont{E.}~\bibnamefont{Accomando}},
  \bibinfo{author}{\bibfnamefont{A.}~\bibnamefont{Denner}}, \bibnamefont{and}
  \bibinfo{author}{\bibfnamefont{S.}~\bibnamefont{Pozzorini}},
  \bibinfo{journal}{Phys. Rev.} \textbf{\bibinfo{volume}{D65}},
  \bibinfo{pages}{073003} (\bibinfo{year}{2002}),
  \eprint{arXiv:hep-ph/0110114v2}.

\bibitem[{\citenamefont{Kuhn et~al.}(2008)\citenamefont{Kuhn, Kulesza,
  Pozzorini, and Schulze}}]{Pozzorini3}
\bibinfo{author}{\bibfnamefont{J.~H.} \bibnamefont{Kuhn}},
  \bibinfo{author}{\bibfnamefont{A.}~\bibnamefont{Kulesza}},
  \bibinfo{author}{\bibfnamefont{S.}~\bibnamefont{Pozzorini}},
  \bibnamefont{and} \bibinfo{author}{\bibfnamefont{M.}~\bibnamefont{Schulze}},
  \bibinfo{journal}{Nucl. Phys.} \textbf{\bibinfo{volume}{B797}},
  \bibinfo{pages}{27} (\bibinfo{year}{2008}), \eprint{arXiv:0708.0476v2}.

\bibitem[{\citenamefont{Fuhrer et~al.}(2011)\citenamefont{Fuhrer, Manohar, and
  Waalewijn}}]{Fuhrer:2010vi}
\bibinfo{author}{\bibfnamefont{A.}~\bibnamefont{Fuhrer}},
  \bibinfo{author}{\bibfnamefont{A.~V.} \bibnamefont{Manohar}},
  \bibnamefont{and} \bibinfo{author}{\bibfnamefont{W.~J.}
  \bibnamefont{Waalewijn}}, \bibinfo{journal}{Phys.Rev.}
  \textbf{\bibinfo{volume}{D84}}, \bibinfo{pages}{013007}
  (\bibinfo{year}{2011}), \eprint{arXiv:1011.1505}.

\bibitem[{\citenamefont{Manohar and Trott}(2012)}]{Manohar:2012}
\bibinfo{author}{\bibfnamefont{A.~V.} \bibnamefont{Manohar}} \bibnamefont{and}
  \bibinfo{author}{\bibfnamefont{M.}~\bibnamefont{Trott}}
  (\bibinfo{year}{2012}), \eprint{arXiv:1201.3926v1}.

\bibitem[{\citenamefont{Cahn et~al.}(1987)\citenamefont{Cahn, Ellis, Kleiss,
  and Stirling}}]{Cahn:1986zv}
\bibinfo{author}{\bibfnamefont{R.~N.} \bibnamefont{Cahn}},
  \bibinfo{author}{\bibfnamefont{S.~D.} \bibnamefont{Ellis}},
  \bibinfo{author}{\bibfnamefont{R.}~\bibnamefont{Kleiss}}, \bibnamefont{and}
  \bibinfo{author}{\bibfnamefont{W.~J.} \bibnamefont{Stirling}},
  \bibinfo{journal}{Phys.Rev.} \textbf{\bibinfo{volume}{D35}},
  \bibinfo{pages}{1626} (\bibinfo{year}{1987}).

\bibitem[{\citenamefont{Barger et~al.}(1995)\citenamefont{Barger, Phillips, and
  Zeppenfeld}}]{Barger:1994zq}
\bibinfo{author}{\bibfnamefont{V.~D.} \bibnamefont{Barger}},
  \bibinfo{author}{\bibfnamefont{R.}~\bibnamefont{Phillips}}, \bibnamefont{and}
  \bibinfo{author}{\bibfnamefont{D.}~\bibnamefont{Zeppenfeld}},
  \bibinfo{journal}{Phys.Lett.} \textbf{\bibinfo{volume}{B346}},
  \bibinfo{pages}{106} (\bibinfo{year}{1995}), \eprint{hep-ph/9412276}.

\bibitem[{\citenamefont{Figy et~al.}(2003)\citenamefont{Figy, Zeppenfeld, and
  Oleari}}]{Figy:2003nv}
\bibinfo{author}{\bibfnamefont{T.}~\bibnamefont{Figy}},
  \bibinfo{author}{\bibfnamefont{D.}~\bibnamefont{Zeppenfeld}},
  \bibnamefont{and} \bibinfo{author}{\bibfnamefont{C.}~\bibnamefont{Oleari}},
  \bibinfo{journal}{Phys. Rev.} \textbf{\bibinfo{volume}{D68}},
  \bibinfo{pages}{073005} (\bibinfo{year}{2003}), \eprint{hep-ph/0306109}.

\bibitem[{\citenamefont{Del~Duca et~al.}(2006)\citenamefont{Del~Duca, Klamke,
  Zeppenfeld, Mangano, Moretti et~al.}}]{DelDuca:2006hk}
\bibinfo{author}{\bibfnamefont{V.}~\bibnamefont{Del~Duca}},
  \bibinfo{author}{\bibfnamefont{G.}~\bibnamefont{Klamke}},
  \bibinfo{author}{\bibfnamefont{D.}~\bibnamefont{Zeppenfeld}},
  \bibinfo{author}{\bibfnamefont{M.~L.} \bibnamefont{Mangano}},
  \bibinfo{author}{\bibfnamefont{M.}~\bibnamefont{Moretti}},
  \bibnamefont{et~al.}, \bibinfo{journal}{JHEP}
  \textbf{\bibinfo{volume}{0610}}, \bibinfo{pages}{016} (\bibinfo{year}{2006}),
  \eprint{hep-ph/0608158}.

\bibitem[{\citenamefont{Ciccolini et~al.}(2007)\citenamefont{Ciccolini, Denner,
  and Dittmaier}}]{Ciccolini:2007jr}
\bibinfo{author}{\bibfnamefont{M.}~\bibnamefont{Ciccolini}},
  \bibinfo{author}{\bibfnamefont{A.}~\bibnamefont{Denner}}, \bibnamefont{and}
  \bibinfo{author}{\bibfnamefont{S.}~\bibnamefont{Dittmaier}},
  \bibinfo{journal}{Phys. Rev. Lett.} \textbf{\bibinfo{volume}{99}},
  \bibinfo{pages}{161803} (\bibinfo{year}{2007}), \eprint{arXiv:0707.0381}.

\bibitem[{\citenamefont{Ciccolini et~al.}(2008)\citenamefont{Ciccolini, Denner,
  and Dittmaier}}]{Ciccolini:2007ec}
\bibinfo{author}{\bibfnamefont{M.}~\bibnamefont{Ciccolini}},
  \bibinfo{author}{\bibfnamefont{A.}~\bibnamefont{Denner}}, \bibnamefont{and}
  \bibinfo{author}{\bibfnamefont{S.}~\bibnamefont{Dittmaier}},
  \bibinfo{journal}{Phys. Rev.} \textbf{\bibinfo{volume}{D77}},
  \bibinfo{pages}{013002} (\bibinfo{year}{2008}), \eprint{arXiv:0710.4749}.

\bibitem[{\citenamefont{Chiu et~al.}(2010)\citenamefont{Chiu, Fuhrer, Kelley,
  and Manohar}}]{Chiu:2009ft}
\bibinfo{author}{\bibfnamefont{J.-y.} \bibnamefont{Chiu}},
  \bibinfo{author}{\bibfnamefont{A.}~\bibnamefont{Fuhrer}},
  \bibinfo{author}{\bibfnamefont{R.}~\bibnamefont{Kelley}}, \bibnamefont{and}
  \bibinfo{author}{\bibfnamefont{A.~V.} \bibnamefont{Manohar}},
  \bibinfo{journal}{Phys. Rev.} \textbf{\bibinfo{volume}{D81}},
  \bibinfo{pages}{014023} (\bibinfo{year}{2010}), \eprint{arXiv:0909.0947}.

\bibitem[{\citenamefont{Han et~al.}(1992)\citenamefont{Han, Valencia, and
  Willenbrock}}]{Han:1992hr}
\bibinfo{author}{\bibfnamefont{T.}~\bibnamefont{Han}},
  \bibinfo{author}{\bibfnamefont{G.}~\bibnamefont{Valencia}}, \bibnamefont{and}
  \bibinfo{author}{\bibfnamefont{S.}~\bibnamefont{Willenbrock}},
  \bibinfo{journal}{Phys.Rev.Lett.} \textbf{\bibinfo{volume}{69}},
  \bibinfo{pages}{3274} (\bibinfo{year}{1992}), \eprint{hep-ph/9206246}.

\bibitem[{\citenamefont{Berger and Campbell}(2004)}]{Berger:2004pca}
\bibinfo{author}{\bibfnamefont{E.~L.} \bibnamefont{Berger}} \bibnamefont{and}
  \bibinfo{author}{\bibfnamefont{J.~M.} \bibnamefont{Campbell}},
  \bibinfo{journal}{Phys.Rev.} \textbf{\bibinfo{volume}{D70}},
  \bibinfo{pages}{073011} (\bibinfo{year}{2004}), \eprint{hep-ph/0403194}.

\bibitem[{\citenamefont{Arnold et~al.}(2009)\citenamefont{Arnold, Bahr, Bozzi,
  Campanario, Englert et~al.}}]{Arnold:2008rz}
\bibinfo{author}{\bibfnamefont{K.}~\bibnamefont{Arnold}},
  \bibinfo{author}{\bibfnamefont{M.}~\bibnamefont{Bahr}},
  \bibinfo{author}{\bibfnamefont{G.}~\bibnamefont{Bozzi}},
  \bibinfo{author}{\bibfnamefont{F.}~\bibnamefont{Campanario}},
  \bibinfo{author}{\bibfnamefont{C.}~\bibnamefont{Englert}},
  \bibnamefont{et~al.}, \bibinfo{journal}{Comput.Phys.Commun.}
  \textbf{\bibinfo{volume}{180}}, \bibinfo{pages}{1661} (\bibinfo{year}{2009}),
  \eprint{arXiv:0811.4559}.

\bibitem[{\citenamefont{Harlander et~al.}(2008)\citenamefont{Harlander,
  Vollinga, and Weber}}]{Harlander:2008xn}
\bibinfo{author}{\bibfnamefont{R.~V.} \bibnamefont{Harlander}},
  \bibinfo{author}{\bibfnamefont{J.}~\bibnamefont{Vollinga}}, \bibnamefont{and}
  \bibinfo{author}{\bibfnamefont{M.~M.} \bibnamefont{Weber}},
  \bibinfo{journal}{Phys.Rev.} \textbf{\bibinfo{volume}{D77}},
  \bibinfo{pages}{053010} (\bibinfo{year}{2008}), \eprint{arXiv:0801.3355}.

\bibitem[{\citenamefont{Bolzoni
  et~al.}(2010{\natexlab{a}})\citenamefont{Bolzoni, Maltoni, Moch, and
  Zaro}}]{Bolzoni:2010xr}
\bibinfo{author}{\bibfnamefont{P.}~\bibnamefont{Bolzoni}},
  \bibinfo{author}{\bibfnamefont{F.}~\bibnamefont{Maltoni}},
  \bibinfo{author}{\bibfnamefont{S.-O.} \bibnamefont{Moch}}, \bibnamefont{and}
  \bibinfo{author}{\bibfnamefont{M.}~\bibnamefont{Zaro}},
  \bibinfo{journal}{Phys. Rev. Lett.} \textbf{\bibinfo{volume}{105}},
  \bibinfo{pages}{011801} (\bibinfo{year}{2010}{\natexlab{a}}),
  \eprint{arXiv:1003.4451}.

\bibitem[{\citenamefont{Bolzoni
  et~al.}(2010{\natexlab{b}})\citenamefont{Bolzoni, Zaro, Maltoni, and
  Moch}}]{Bolzoni:2010as}
\bibinfo{author}{\bibfnamefont{P.}~\bibnamefont{Bolzoni}},
  \bibinfo{author}{\bibfnamefont{M.}~\bibnamefont{Zaro}},
  \bibinfo{author}{\bibfnamefont{F.}~\bibnamefont{Maltoni}}, \bibnamefont{and}
  \bibinfo{author}{\bibfnamefont{S.-O.} \bibnamefont{Moch}},
  \bibinfo{journal}{Nucl. Phys. Proc. Suppl.}
  \textbf{\bibinfo{volume}{205-206}}, \bibinfo{pages}{314}
  (\bibinfo{year}{2010}{\natexlab{b}}), \eprint{arXiv:1006.2323}.

\bibitem[{\citenamefont{Bolzoni et~al.}(2012)\citenamefont{Bolzoni, Maltoni,
  Moch, and Zaro}}]{Bolzoni:2011cu}
\bibinfo{author}{\bibfnamefont{P.}~\bibnamefont{Bolzoni}},
  \bibinfo{author}{\bibfnamefont{F.}~\bibnamefont{Maltoni}},
  \bibinfo{author}{\bibfnamefont{S.-O.} \bibnamefont{Moch}}, \bibnamefont{and}
  \bibinfo{author}{\bibfnamefont{M.}~\bibnamefont{Zaro}},
  \bibinfo{journal}{Phys.Rev.} \textbf{\bibinfo{volume}{D85}},
  \bibinfo{pages}{035002} (\bibinfo{year}{2012}), \eprint{arXiv:1109.3717}.

\bibitem[{\citenamefont{Siringo}(2004{\natexlab{a}})}]{Siringo:2004}
\bibinfo{author}{\bibfnamefont{F.}~\bibnamefont{Siringo}},
  \bibinfo{journal}{Eur. Phys. J.} \textbf{\bibinfo{volume}{C32}},
  \bibinfo{pages}{555} (\bibinfo{year}{2004}{\natexlab{a}}),
  \eprint{arXiv:hep-ph/0307320v1}.

\bibitem[{\citenamefont{Siringo}(2004{\natexlab{b}})}]{Siringo:2004L}
\bibinfo{author}{\bibfnamefont{F.}~\bibnamefont{Siringo}},
  \bibinfo{journal}{Phys. Rev. Lett.} \textbf{\bibinfo{volume}{92}},
  \bibinfo{pages}{119101} (\bibinfo{year}{2004}{\natexlab{b}}).

\bibitem[{\citenamefont{Siringo and Marotta}(2006)}]{Siringo:2006}
\bibinfo{author}{\bibfnamefont{F.}~\bibnamefont{Siringo}} \bibnamefont{and}
  \bibinfo{author}{\bibfnamefont{L.}~\bibnamefont{Marotta}},
  \bibinfo{journal}{Phys. Rev.} \textbf{\bibinfo{volume}{D74}},
  \bibinfo{pages}{115001} (\bibinfo{year}{2006}),
  \eprint{arXiv:hep-ph/0605276v2}.

\bibitem[{\citenamefont{Korchemsky and Radyushkin}(1987)}]{Korchemsky:1987wg}
\bibinfo{author}{\bibfnamefont{G.}~\bibnamefont{Korchemsky}} \bibnamefont{and}
  \bibinfo{author}{\bibfnamefont{A.}~\bibnamefont{Radyushkin}},
  \bibinfo{journal}{Nucl.Phys.} \textbf{\bibinfo{volume}{B283}},
  \bibinfo{pages}{342} (\bibinfo{year}{1987}).

\bibitem[{\citenamefont{Moch et~al.}(2004)\citenamefont{Moch, Vermaseren, and
  Vogt}}]{Moch:2004pa}
\bibinfo{author}{\bibfnamefont{S.}~\bibnamefont{Moch}},
  \bibinfo{author}{\bibfnamefont{J.}~\bibnamefont{Vermaseren}},
  \bibnamefont{and} \bibinfo{author}{\bibfnamefont{A.}~\bibnamefont{Vogt}},
  \bibinfo{journal}{Nucl.Phys.} \textbf{\bibinfo{volume}{B688}},
  \bibinfo{pages}{101} (\bibinfo{year}{2004}), \eprint{hep-ph/0403192}.

\bibitem[{\citenamefont{Chay and Kim}(2002)}]{Chay:2002vy}
\bibinfo{author}{\bibfnamefont{J.}~\bibnamefont{Chay}} \bibnamefont{and}
  \bibinfo{author}{\bibfnamefont{C.}~\bibnamefont{Kim}},
  \bibinfo{journal}{Phys.Rev.} \textbf{\bibinfo{volume}{D65}},
  \bibinfo{pages}{114016} (\bibinfo{year}{2002}), \eprint{hep-ph/0201197}.

\bibitem[{\citenamefont{Manohar et~al.}(2002)\citenamefont{Manohar, Mehen,
  Pirjol, and Stewart}}]{Manohar:2002fd}
\bibinfo{author}{\bibfnamefont{A.~V.} \bibnamefont{Manohar}},
  \bibinfo{author}{\bibfnamefont{T.}~\bibnamefont{Mehen}},
  \bibinfo{author}{\bibfnamefont{D.}~\bibnamefont{Pirjol}}, \bibnamefont{and}
  \bibinfo{author}{\bibfnamefont{I.~W.} \bibnamefont{Stewart}},
  \bibinfo{journal}{Phys. Lett.} \textbf{\bibinfo{volume}{B539}},
  \bibinfo{pages}{59} (\bibinfo{year}{2002}), \eprint{hep-ph/0204229}.

\bibitem[{\citenamefont{Stewart et~al.}(2010)\citenamefont{Stewart, Tackmann,
  and Waalewijn}}]{Stewart:2010tn}
\bibinfo{author}{\bibfnamefont{I.~W.} \bibnamefont{Stewart}},
  \bibinfo{author}{\bibfnamefont{F.~J.} \bibnamefont{Tackmann}},
  \bibnamefont{and} \bibinfo{author}{\bibfnamefont{W.~J.}
  \bibnamefont{Waalewijn}}, \bibinfo{journal}{Phys.Rev.Lett.}
  \textbf{\bibinfo{volume}{105}}, \bibinfo{pages}{092002}
  (\bibinfo{year}{2010}), \eprint{arXiv:1004.2489}.

\bibitem[{\citenamefont{Eidelman et~al. [Particle Data~Group]}(2004)}]{PDG}
\bibinfo{author}{\bibfnamefont{S.}~\bibnamefont{Eidelman et~al. [Particle
  Data~Group]}}, \bibinfo{journal}{Phys. Lett.}
  \textbf{\bibinfo{volume}{B592}}, \bibinfo{pages}{1} (\bibinfo{year}{2004}).

\bibitem[{\citenamefont{Arason et~al.}(1992)\citenamefont{Arason, Castano,
  Kesthelyi, Mikaelian, Piard, Ramond, and Wright}}]{Arason}
\bibinfo{author}{\bibfnamefont{H.}~\bibnamefont{Arason}},
  \bibinfo{author}{\bibfnamefont{D.~J.} \bibnamefont{Castano}},
  \bibinfo{author}{\bibfnamefont{B.}~\bibnamefont{Kesthelyi}},
  \bibinfo{author}{\bibfnamefont{S.}~\bibnamefont{Mikaelian}},
  \bibinfo{author}{\bibfnamefont{E.~J.} \bibnamefont{Piard}},
  \bibinfo{author}{\bibfnamefont{P.}~\bibnamefont{Ramond}}, \bibnamefont{and}
  \bibinfo{author}{\bibfnamefont{B.~D.} \bibnamefont{Wright}},
  \bibinfo{journal}{Phys. Rev. D} \textbf{\bibinfo{volume}{46}},
  \bibinfo{pages}{3945} (\bibinfo{year}{1992}).

\bibitem[{\citenamefont{Siringo and Marotta}(2008)}]{Siringo:2008}
\bibinfo{author}{\bibfnamefont{F.}~\bibnamefont{Siringo}} \bibnamefont{and}
  \bibinfo{author}{\bibfnamefont{L.}~\bibnamefont{Marotta}},
  \bibinfo{journal}{Phys. Rev.} \textbf{\bibinfo{volume}{D78}},
  \bibinfo{pages}{016003} (\bibinfo{year}{2008}), \eprint{arXiv:0803.3043}.

\bibitem[{\citenamefont{Siringo and Marotta}(2010)}]{Siringo:2010}
\bibinfo{author}{\bibfnamefont{F.}~\bibnamefont{Siringo}} \bibnamefont{and}
  \bibinfo{author}{\bibfnamefont{L.}~\bibnamefont{Marotta}},
  \bibinfo{journal}{Int. J. Mod. Phys.} \textbf{\bibinfo{volume}{A25}},
  \bibinfo{pages}{5865} (\bibinfo{year}{2010}), \eprint{arXiv:0901.2418}.

\bibitem[{\citenamefont{Marotta and Siringo}(2012)}]{Siringo:2012}
\bibinfo{author}{\bibfnamefont{L.}~\bibnamefont{Marotta}} \bibnamefont{and}
  \bibinfo{author}{\bibfnamefont{F.}~\bibnamefont{Siringo}},
  \bibinfo{journal}{Modern Physics Letters} \textbf{\bibinfo{volume}{B26}},
  \bibinfo{pages}{1250130} (\bibinfo{year}{2012}), \eprint{arXiv:0806.4569v3}.

\bibitem[{\citenamefont{Siringo and Marotta}(2005)}]{Siringo:2005}
\bibinfo{author}{\bibfnamefont{F.}~\bibnamefont{Siringo}} \bibnamefont{and}
  \bibinfo{author}{\bibfnamefont{L.}~\bibnamefont{Marotta}},
  \bibinfo{journal}{Eur. Phys. J.} \textbf{\bibinfo{volume}{C44}},
  \bibinfo{pages}{293} (\bibinfo{year}{2005}), \eprint{arXiv:hep-ph/0506284}.

\bibitem[{\citenamefont{Pumplin et~al.}(2006)\citenamefont{Pumplin, Belyaev,
  Huston, Stump, and Tung}}]{PDF}
\bibinfo{author}{\bibfnamefont{J.}~\bibnamefont{Pumplin}},
  \bibinfo{author}{\bibfnamefont{A.}~\bibnamefont{Belyaev}},
  \bibinfo{author}{\bibfnamefont{J.}~\bibnamefont{Huston}},
  \bibinfo{author}{\bibfnamefont{D.}~\bibnamefont{Stump}}, \bibnamefont{and}
  \bibinfo{author}{\bibfnamefont{W.~K.} \bibnamefont{Tung}},
  \bibinfo{journal}{JHEP} \textbf{\bibinfo{volume}{0602}}, \bibinfo{pages}{032}
  (\bibinfo{year}{2006}), \eprint{hep-ph/0512167}.

\bibitem[{\citenamefont{Siringo}(2000)}]{Siringo:2000}
\bibinfo{author}{\bibfnamefont{F.}~\bibnamefont{Siringo}},
  \bibinfo{journal}{Phys. Rev.} \textbf{\bibinfo{volume}{D62}},
  \bibinfo{pages}{116009} (\bibinfo{year}{2000}),
  \eprint{arXiv:hep-ph/0008030}.

\bibitem[{\citenamefont{Siringo}(2002)}]{Siringo:2002}
\bibinfo{author}{\bibfnamefont{F.}~\bibnamefont{Siringo}},
  \bibinfo{journal}{Europhys. Lett.} \textbf{\bibinfo{volume}{59}},
  \bibinfo{pages}{820} (\bibinfo{year}{2002}), \eprint{arXiv:hep-ph/0105018}.

\bibitem[{\citenamefont{Passarino and Veltman}(1979)}]{Passarino:1978jh}
\bibinfo{author}{\bibfnamefont{G.}~\bibnamefont{Passarino}} \bibnamefont{and}
  \bibinfo{author}{\bibfnamefont{M.}~\bibnamefont{Veltman}},
  \bibinfo{journal}{Nucl.Phys.} \textbf{\bibinfo{volume}{B160}},
  \bibinfo{pages}{151} (\bibinfo{year}{1979}).

\bibitem[{\citenamefont{Fleischer and Jegerlehner}(1981)}]{Fleischer:1980ub}
\bibinfo{author}{\bibfnamefont{J.}~\bibnamefont{Fleischer}} \bibnamefont{and}
  \bibinfo{author}{\bibfnamefont{F.}~\bibnamefont{Jegerlehner}},
  \bibinfo{journal}{Phys.Rev.} \textbf{\bibinfo{volume}{D23}},
  \bibinfo{pages}{2001} (\bibinfo{year}{1981}).

\end{thebibliography}

\end{document}